%% file: main.tex
\documentclass[11pt]{article}

\usepackage{graphicx,amssymb,amsmath,amsthm,amsfonts}
\usepackage[font=small,labelfont=bf]{caption}
\usepackage[margin=1.in]{geometry}
\usepackage{subcaption}
\usepackage{comment}

\title{%Covert Computation: Memoryless Nanoscale Cryptography and Biomedical Privacy \\
Covert Computation in Self-Assembled Circuits \thanks{This research was supported in part by National Science Foundation Grant CCF-1817602.}}

\author{Angel A. Cantu \and Austin Luchsinger \and Robert Schweller \and Tim Wylie
}
\index{}
\date{}

\newtheorem{definition}{Definition}
\newtheorem{theorem}{Theorem}

\newtheorem{conjecture}{Conjecture}

\begin{document}
\thispagestyle{empty}
\maketitle
\begin{center}
\vspace{-.5cm}
\small
Department of Computer Science \\ University of Texas - Rio Grande Valley \\ {\tt \{angel.cantu01, austin.luchsinger01, robert.schweller, timothy.wylie\}@utrgv.edu}
\end{center}

\begin{abstract}
  Traditionally, computation within self-assembly models is hard to conceal because the self-assembly process generates a crystalline assembly whose computational history is inherently part of the structure itself.  With no way to remove information from the computation, this computational model offers a unique problem:  how can computational input and computation be hidden while still computing and reporting the final output? Designing such systems is inherently motivated by privacy concerns in biomedical computing and applications in cryptography.

  In this paper we propose the problem of performing ``covert computation'' within tile self-assembly that seeks to design self-assembly systems that ``conceal'' both the input and computational history of performed computations.  We achieve these results within the growth-only restricted abstract tile assembly model (aTAM) with positive and negative interactions.  We show that general-case covert computation is possible by implementing a set of basic covert logic gates capable of simulating any circuit (functionally complete).  To further motivate the study of covert computation, we apply our new framework to resolve an outstanding complexity question; we use our covert circuitry to show that the unique assembly verification problem within the growth-only aTAM with negative interactions is coNP-complete.
\end{abstract}
\thispagestyle{empty}

\setcounter{page}{1}
\input{robbieintro}

\input{definitions}

\input{uavgadgets}

\input{uav}

\input{examples}

\input{conclusion}

%\section{Acknowledgements}
%We would like to thank the anonymous reviewers for their careful review of our work and for their constructive feedback.

%bibliography
\bibliographystyle{plain}
\bibliography{covert}

%\newpage
%\appendix
%\input{properties}

\end{document}

%% file: robbieintro.tex
\section{Introduction} \label{sec:intro}
%for robbie to write

%\section{The REAL Introduction}
%for tim to write

%\section{The For REALZ Introduction}
%for Robbie to write
%
%
%
%
%reasons robbie should write this introduction:
%1. Tim called "not it" first.
%2. "not it"s are legally binding.
%3. Robbie is probably a spy
%4. He left town the day we started writing this paper.
%
%
%motivations
%1. theoretical uav, etc. in physical computations
%2. biomedical privacy
%3. military applications
%
%Also, we did some growth-only stuff already \cite{Chalk:2018:ESA}.

%\paragraph{Algorithmic Self-Assembly Background}
%\begin{itemize}
%    \item self-assembly
%    \item tile assembly model
%    \item it's a computational model
%\end{itemize}

Since the discovery of DNA over half a century ago, humans have been continually working to understand and harness the vast amount of information it contains. The Human Genome Project \cite{Lander:2001:NATURE}, which began in 1990 and took a decade, was the first major attempt to fully sequence the human genome. In the years since, sequencing has become extremely cheap and easy, and our ability to manipulate DNA has emerged as a central tool for many applications related to nanotechnology and biomedical engineering.

Although this progress has many benefits, as we learn more about the information, we also must be careful with the shared data. There are databases of anonymous DNA sequences, which can sometimes be deanonymized with only small amounts of information such as a surname \cite{Gymrek:2013:SCI}, or by reconstructing physical features from the DNA \cite{Claes:2014:PLOSG}.
In order to address these issues, there has been work on cryptographic schemes aimed at obscuring results related to DNA or the input/output \cite{DeCristofaro:2013:WPES,Dowlin:2017:PIEEE,Huang:2015:SSP,Yang:2014:CSB}.

In this work we take the first steps in addressing some of these issues within self-assembling systems by proposing a new style of computation termed \emph{covert computation} with important motivations for private biomedical computing and cryptography.
%\textbf{Self-Assembly.}
%Another fundamental aspect in the rise of nanotechnology and molecular computation has been the use of self-assembly systems based on DNA and learning how to design and program these systems.
Self-assembly is the process by which systems of simple objects autonomously organize themselves through local interactions into larger, more complex objects. %Self-assembly processes are abundant in nature and serve as the basis for biological growth and replication.
Understanding how to design and efficiently program molecular self-assembly systems is fundamental for the future of nanotechnology.
The abstract Tile Self-Assembly Model (aTAM)~\cite{DotCACM,Patitz2014}, motivated by a DNA implementaiton~\cite{ContantineThesis}, has become the premiere model for the study of the computational power of self-assembling systems.  In the aTAM, system monomers are modeled by four-sided Wang tiles which randomly combine and attach if the respective bonding domains on tile edges are sufficiently strong.  The aTAM is known to be computationally universal~\cite{Winf98} and intrinsically universal~\cite{USAreal}.  %In this paper, we propose a new style of computation termed \emph{covert computation} with important motivations for private biomedical computing and cryptography.

\textbf{Covert Computation.}
As a computational model, tile self-assembly differs from  traditional models of computation in that computational steps are defined by permanently placing particular tile types at specific locations in geometric space.  A history of each computational step is thereby recorded in the final assembled structure.  This presents a unique problem to this type of computation:  is it possible to conceal the input and history of a computation within the final assembly while still computing and reporting the output of the computation?
%Consider an $n$-bit input decision problem which has exponentially many distinct inputs and computational histories, but only two outputs.
%In this scenario, the goal of a \emph{covert} solution is a system that assembles only two output assemblies, one for each output, regardless of the exponentially many possible inputs.
Concealing the computational histories of the self-assembly process in this way requires designing a computational system which encodes computational steps in the \emph{order} of tile placement, rather than the type and location of tile placements. We use the term \emph{covert}\footnote{It is important to note that the term \emph{covert} has specific meaning in cryptography which does not apply here.} to describe this concealment of inputs and computational histories. This method of computing is different than previous tile self-assembly computing methods and requires novel techniques.

Also, while the reader may notice many parallels between our work and traditional secure multiparty computation \cite{chaum1988}, it should be made clear that our main result is the secure computation of a function with only a single party. The challenges presented above make this an interesting problem for tile self-assembly.

\textbf{Motivation.}
%\begin{itemize}
%    \item Biomedical privacy
%    \item Cryptography
%    \item Military application
%    \item Complexity applications such as UAV
%\end{itemize}
The concept of covert computation within self-assembly has many potential applications.  We briefly outline a few biomedical computing applications.  %Established DNA based implementations of tile self-assembly make the interface of tile assembly systems with biology-based inputs quite feasible.
%One potential application would be the design of a set of diagnostic tiles to which a patient adds a biological input such as a blood sample.  From this the diagnostic system could compute some desired function that outputs specific diagnostic statistics.  In a covert computing system, the patient could be sent the diagnostic system as a droplet of DNA, and personally combine their biological sample.  The combined product could then be sent to a medical facility for interpretation, thereby ensuring privacy based on the covert computation performed, i.e., only the results can be read by the lab and the user's DNA input is obscured.
Consider a set of diagnostic tiles sent to a patient as a droplet of DNA to which the patient adds some biological input such as a blood sample.  From this the diagnostic system could compute some desired function that outputs specific diagnostic statistics. The patient sends the combined product to a medical facility for interpretation. With covert computation, only the results can be read by the lab and the user's biological input is obscured ensuring privacy. %In a covert computing system, the patient could be sent the diagnostic system as a droplet of DNA, add their biological sample, and send the combined product to a medical facility for interpretation. With covert computation, only the results can be read by the lab and the user's biological input is obscured ensuring privacy.

Another potential use involves implementing a cryptography system within a molecular computing framework.  The ability to covertly compute allows users to provide a personal key input that may be combined with a publicly available covert system where the combination  verifies some computable property of the input key without revealing any additional details of the key.  This style of cryptographic scheme fits well when the input keys are biological based inputs.

A final potential biological application might be engineering a system for unlocking key biological properties within bio-engineered crops.  For example, by releasing a hidden ``key'' input, covert computation might allow a field of crops to become fertile.  A company owning the patent on this type of activation might desire the security of ensuring that the release key cannot be deciphered from the activated crop based on a covert molecular computation.

The final motivation of covert computation is within algorithmic self-assembly.  We believe the concept of covert computation is fundamental and hope that our novel design techniques will be applicable to a number of future problems in the area.  As evidence towards this, we apply our techniques to resolve the complexity of the fundamental question of verifying whether a tile system uniquely assembles a given assembly within the growth-only negative-glue aTAM.

\textbf{Contributions.}
After formally defining the concept of covert computation in tile self-assembly, we implement several covert logic gates within the negative-glue growth-only abstract Tile Assembly Model (this growth-only restriction to negative glues has been seen in the 2HAM \cite{Chalk:2018:ESA}, and negative glues in tile assembly  have received extensive study~\cite{USR2017SODA,Doty2013,LSW2017SAS,rgTAM,PRS2016RMN,REIF20111592,SS2013FEC}), and show these gates may be combined to create general circuits, thereby showing that general covert computation is possible.  Finally, we apply our techniques and framework to address the fundamental problem of deciding if a negative-glue aTAM system uniquely produces a given assembly.  We show this problem is coNP-complete. Table \ref{tab:uav} outlines how our result compares to what was previously known about Unique Assembly Verification in the aTAM.

\begin{table}
    \begin{center}
        \begin{tabular}{ | c | c | c | c | c |}
            \hline
            \textbf{Model} & \textbf{Negative Glues} & \textbf{Detachment} & \textbf{Complexity} & \textbf{Theorem} \\ \hline
            aTAM & No & No & $O(|A|^2+|A||T|)$ & Thm. 3.2 in \cite{ACGHKMR02} \\ \hline
            aTAM & Yes & No & coNP-complete & Thm. \ref{thm:uav}  \\ \hline
            aTAM & Yes & Yes & Undecidable & \cite{Doty2013} \\ \hline
        \end{tabular}
        \vspace*{.2cm}
        \caption{The complexity of Unique Assembly Verification in the aTAM in relation to negative glues. $|A|$ refers to the size of an assembly and $|T|$ is the number of tile types.}\label{tab:uav}
    \end{center}
\end{table}

This work is an extension of a paper originally published in \cite{Cantu:2019:ICALP}. We have included a lot more detail as well as additional examples and gadgets. There was also a small issue with the NAND gadget related to backfilling which has been corrected. The backfill stop gadget was also added, which allows backfilling to begin along a wire for a cleaner construction with the FANOUT and at the output.

%\subsection{Related Work}
%
%\begin{itemize}
%    \item Negative glue papers
%    \item growth only paper(s)
%    \item computation specific papers
%    \item circuit building papers
%\end{itemize}

%% file: definitions.tex
%\vspace*{-.2cm}
\section{Definitions}\label{sec:def}
%\vspace*{-.2cm}

%
%to define:
%tile/atam/etc.
%-negative glues
%-growth only system
%-covert computation
%-input/output
%-gates
%-uav
%-backfill/backpropogate
%-universal computation? only need a nand gate, crossover, and fanout
%--maybe standard symbols

%There are several preliminaries before the main results.
We begin with an overview of the abstract Tile-Assembly Model (aTAM) and then give the new definitions introducing covert computation. Due to the extensive use of the aTAM in the literature, we only give a high-level overview of the aTAM. %Formal definitions for all the concepts are in Appendix \ref{app:def}.

%\vspace*{-.3cm}
\subsection{Abstract Tile Assembly Model}
%\vspace*{-.2cm}

\begin{figure}[t!]
    %\vspace*{-.3cm}
    \centering
    % \begin{subfigure}[b]{0.31\textwidth}
    %     \includegraphics[width=1.\textwidth]{images/atam.png}
    %     \caption{aTAM}
    %     \label{fig:atam}
    % \end{subfigure}
    % \hspace*{.2cm}
    \begin{subfigure}[b]{0.49\textwidth}
        \includegraphics[width=1.\textwidth]{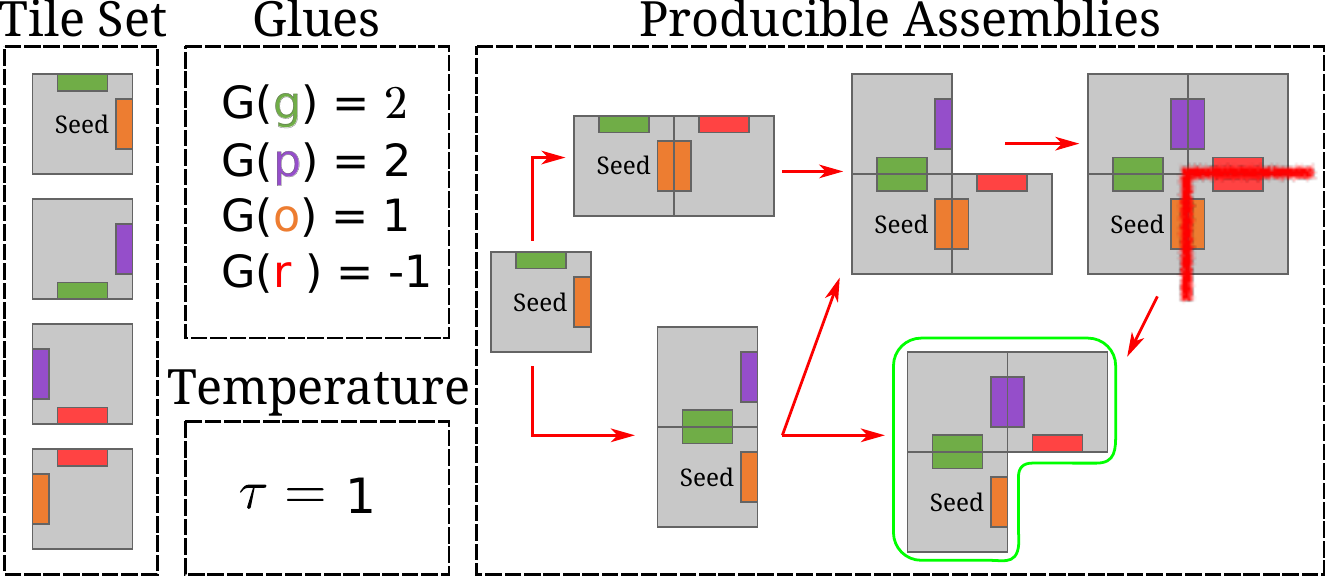}
        \caption{Negative aTAM}
        \label{fig:negatam}
    \end{subfigure}
    \hspace*{.5cm}
    \begin{subfigure}[b]{0.42\textwidth}
        \includegraphics[width=1.\textwidth]{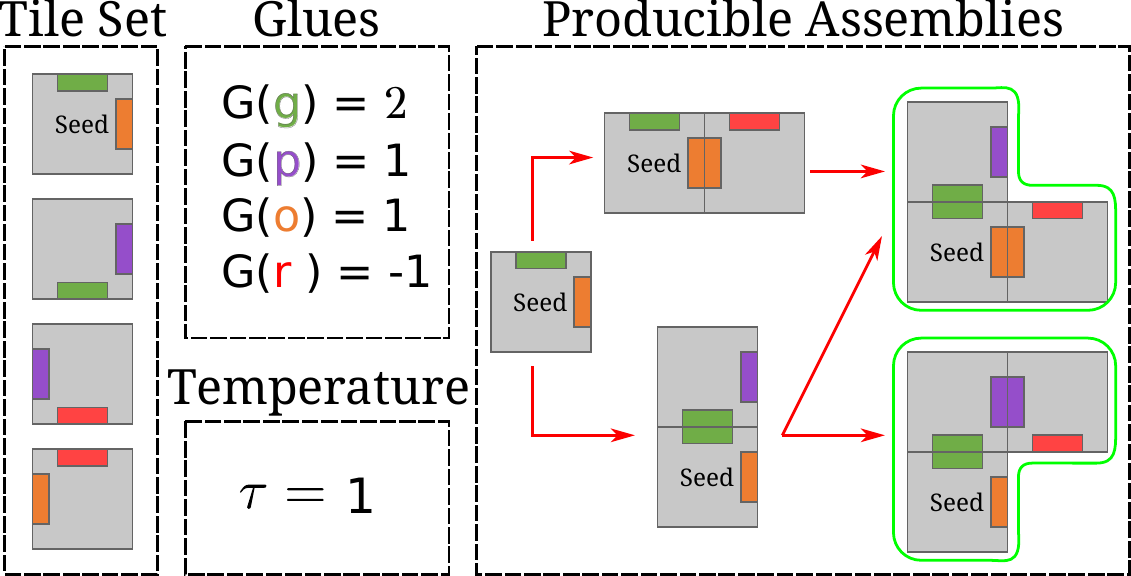}
        \caption{Growth aTAM}
        \label{fig:growthatam}
    \end{subfigure}
    %\vspace*{-.2cm}
    \caption{High-level overview of the aTAM with repulsive forces. Both systems have tiles that can attach to the seed tile given they can attach with $\tau$ strength. The arrows show the possible assembly paths from the seed tile with the terminal assembly being outlined.
    (a) A negative aTAM system that has a possible assembly path causing disassembly. One path is growth-only, but the other path can attach the tile with the purple/red glues, which causes the orange/red tile to become unstable and detach. (b) A growth-only aTAM system where negative glues are used to block, but never cause disassembly. The only difference is that the purple glue attaches with strength 1, $G(p)=1$. This yields two possible terminal assemblies, neither of which include disassembly.}
    \label{fig:models}
    %\vspace*{-.5cm}
\end{figure}

Figure \ref{fig:models} gives a high-level overview of the models with a couple of example systems. Essentially, we have non-rotating square \emph{tiles} that have a \emph{glue} label on each edge. The tile with its labels is a \emph{tile type}. The \emph{tile set} is all the tile types. A glue function determines the strength of matching glue labels. An \emph{assembly} is a single tile or a finite set of tiles that have combined via the glues.
If the combined strength of the glue labels of a single attaching tile to an assembly is greater than or equal to the \emph{temperature} $\tau$, the tile may attach. A \emph{producible} assembly is any assembly that might be achieved by beginning with the \emph{seed} (the specified starting assembly) and attaching tiles.  A producible assembly is further said to be \emph{terminal} if no further tile attachment is possible.  A tile system is said to \emph{uniquely produce} a (terminal) assembly $A$ if all producible assemblies will eventually grow into $A$.  A tile system is formally represented as an ordered triplet $\Gamma = (T,S, \tau)$ representing the tile set, seed assembly, and temperature parameter of the system respectively.

%Tiles, tile sets, assemblies, complexity, glues, temperature, seed, producibles, valid, then growth only

In a standard aTAM system, all glues are positive integral values, but here we look at the negative aTAM where the glues may be negative/repulsive.  Such repulsive forces may be used to block the attachment of tiles despite the presence of strong attractive glues.  Moreover, the inclusion of repulsive forces may yield unstable producible assemblies where a subassembly could detach because it no longer has enough binding strength.  While this type of detachment has been studied in the literature~\cite{Doty2013,SS2013FEC}, we avoid this feature in this work as its inclusion drastically changes the complexity of the model by making most types of verification problems undecidable, and may require more sophisticated techniques for experimental implementation.  Thus, we consider a system to be a \emph{valid growth-only} system if all producible assemblies are $\tau$-stable. %The existence of negative strength glues allows for the possibility that unstable assemblies are produced. %Unique Assembly Verification is undecidable when unstable assemblies are allowed.
In this paper we restrict our consideration to valid growth-only systems.

%%%%%%%%%%%%%%%%%%%%%%%%%%%%%%%%%%%%%%%%%%%%%%%%%%%%%%%%%%%%%%%%%%%%%%%%%%%5555

\subsection{Formal Definitions} \label{app:def}

\textbf{Tiles.}
Let $\Pi$ be an alphabet of symbols called the \emph{glue types}.  A tile is a finite edge polygon with some finite subset of border points each assigned a glue type from $\Pi$.  Each glue type $g \in \Pi$ also has some integer strength $str(g)$.
Here, we consider unit square tiles of the same orientation with at most one glue type per face, and the \emph{location} to be the center of the  tile  located at integer coordinates.

\textbf{Assemblies.}
An assembly $A$ is a finite set of tiles whose interiors do not overlap.  %Further, to simplify formalization in this paper, we require the center of each tile in an assembly to be an integer coordinate.
If each tile in $A$ is a translation of some tile in a set of tiles $T$, we say that $A$ is an assembly over tile set $T$.   For a given assembly $A$, define the \emph{bond graph} $G_A$ to be the weighted graph in which each element of $A$ is a vertex, and the weight of an edge between two tiles is the strength of the overlapping matching glue points between the two tiles.  Only overlapping glues of the same type contribute a non-zero weight, whereas overlapping, non-equal glues  contribute zero weight to the bond graph.  The property that only equal glue types interact with each other is referred to as the \emph{diagonal glue function} property and is perhaps more feasible than more general glue functions for experimental implementation (see~\cite{AGKS05g} for the theoretical impact of relaxing this constraint).  An assembly $A$ is said to be \emph{$\tau$-stable} for an integer $\tau$ if the min-cut of $G_A$ has weight at least $\tau$.

\textbf{Tile Attachment.}
Given a tile $t$, an integer $\tau$, and an assembly $A$, we say that $t$ may attach to $A$ at temperature $\tau$ to form $A'$ if there exists a translation $t'$ of $t$ such that $A' = A \cup \{t'\}$, and the sum of newly bonded glues between $t'$ and $A$ meets or exceeds $\tau$.  For a tile set $T$ we use notation $A \rightarrow_{T,\tau} A'$ to denote  there exists some $t\in T$ that may attach to $A$ to form $A'$ at temperature $\tau$.  When $T$ and $\tau$ are implied, we simply say $A \rightarrow A'$.  Further, we say that $A \rightarrow^*A'$ if either $A=A'$, or there exists a finite sequence of assemblies $\langle A_1 \ldots A_k\rangle$ such that $A \rightarrow A_1 \rightarrow \ldots \rightarrow A_k \rightarrow A'$.

\textbf{Tile Systems.}
A tile system $\Gamma = (T,S, \tau)$ is an ordered triplet consisting of a set of tiles $T$ called the system's \emph{tile set}, a $\tau$-stable assembly $S$ called the system's \emph{seed} assembly, and a positive integer $\tau$ referred to as the system's \emph{temperature}. A tile system $\Gamma = (T,S,\tau)$ has an associated set of \emph{producible} assemblies, $\texttt{PROD}_\Gamma$, which define what assemblies can grow from the initial seed $S$ by any sequence of temperature $\tau$ tile attachments from $T$.  Formally, $S \in \texttt{PROD}_\Gamma$ is a base case producible assembly.  Further, for every $A\in \texttt{PROD}_\Gamma$, if $A \rightarrow_{T,\tau} A'$, then $A' \in \texttt{PROD}_\Gamma$.  That is, assembly $S$ is producible, and for every producible assembly $A$, if $A$ can grow into $A'$, then $A'$ is also producible.  %In other words, the set of producible assemblies contains any assembly that can be obtained by any valid sequence of tile attachments starting from assembly $S$.
We further denote a producible assembly $A$ to be \emph{terminal} if $A$ has no attachable tile from $T$ at temperature $\tau$.  We say a system $\Gamma=(T,S,\tau)$ \emph{uniquely produces} an assembly $A$ if all producible assemblies can grow into $A$ through some sequence of tile attachments.  More formally, $\Gamma$ \emph{uniquely produces} an assembly $A \in \texttt{PROD}_\Gamma$ if for every $A' \in \texttt{PROD}_\Gamma$ it is the case that $A' \rightarrow^* A$.  Systems that uniquely produce one assembly are said to be \emph{deterministic}. 

Finally, we consider a system to be a \emph{valid growth-only} system if all assemblies in $\texttt{PROD}_\Gamma$ are $\tau$-stable. The existence of negative strength glues allows for the possibility that unstable assemblies are produced. %Unique Assembly Verification is undecidable when unstable assemblies are allowed. 
%Thus, in this paper we restrict our consideration to valid growth-only systems.

%%%%%%%%%%%%%%%%%%%%%%%%%%%%%%%%%%%%%%%%%%%%%%%%%%%%%%%%%%%%%%%%%%%%%%%%%%%%%%%%55
%\vspace*{-.2cm}
\subsection{Covert Computation}

Here, we provide  formal definitions for computing a function with a tile system, and the further requirement for covert computation of a function.  Our formulation of computing functions is based on that of~\cite{Keenan2016} but modified to allow for each bit to be represented by a sub-assembly potentially larger than a single tile.
%Since this is in a growth only model, we use dual-rail logic.

Informally, a Tile Assembly Computer (TAC) for a function $f$ consists of a set of tiles, along with a format for both input and output.  The input format is a specification for how to build an input seed for the system that encodes the desired input bit-string for function $f$.  We require that each bit of the input be mapped to one of two assemblies for the respective bit position:  a sub-assembly representing ``0'', or a sub-assembly representing ``1''.  The input seed for the entire string is the union of all these sub-assemblies.  This seed, along with the tile set of the TAC, forms a tile system.  The output of the computation is the final terminal assembly this system builds.  To interpret what bit-string is represented by the output, a second \emph{output} format specifies a pair of sub-assemblies for each bit.  The bit-string represented by the union of these subassemblies within the constructed assembly is the output of the system.

For a TAC to \emph{covertly} compute  $f$, the TAC must compute $f$ and produce a unique assembly for each possible output of $f$.  We note that our formulation for providing input and interpreting output is quite rigid and may prohibit more exotic forms of computation.  We acknowledge this, but caution that any formulation must take care to prevent ``cheating'' that could allow the output of a function to be partially or completely encoded within the input, for example.  To prevent this, some type of \emph{uniformity} constraint, similar to what is considered in circuit complexity~\cite{Vollmer:1999:ICC:520668}, should be enforced.
We now provide the formal definitions of function computing and covert computation.

\textbf{Input/Output Templates.}
An $n$-bit input/output template over tile set $T$ is a sequence of ordered pairs of assemblies over $T$: $A= (A_{0,0}, A_{0,1}),$ $\ldots , (A_{n-1, 0},$ $A_{n-1,1})$.  For a given $n$-bit string $b=b_0,\ldots,b_{n-1}$ and $n$-bit input/output template $A$, the \emph{representation} of $b$ with respect to $A$ is the assembly $A(b) = \bigcup_i A_{i,b_i}$.  A template is valid for a temperature $\tau$ if this union never contains overlaps for any choice of $b$, and is always $\tau$-stable.  An assembly $B \supseteq A(b)$, which contains $A(b)$ as a subassembly, is said to represent $b$ as long as $A(d) \nsubseteq B$ for any $d \neq b$.

\textbf{Function Computing Problem.}
A \emph{tile assembly computer} (TAC) is an ordered quadruple $\Im=(T, I, O, \tau)$ where $T$ is a tile set, $I$ is an $n$-bit input template, and $O$ is a $k$-bit output template.  A TAC is said to compute function $f: \mathbb{Z}_{2}^n \rightarrow \mathbb{Z}_{2}^k$ if for any $b \in \mathbb{Z}_{2}^n$ and $c \in \mathbb{Z}_{2}^k$ such that $f(b)=c$, then the tile system $\Gamma_{\Im,b} = (T, I(b), \tau)$ uniquely assembles a set of assemblies at temperature $\tau$ which all represent $c$ with respect to template $O$.

\textbf{Covert Computation.}
A TAC  \emph{covertly} computes a function $f(b)=c$ if 1) it computes $f$, and 2) for each $c$, there exists a unique assembly $A_c$ such that for all $b$, where $f(b)=c$, the system $\Gamma_{\Im,b} = (T, I(b),\tau)$ uniquely produces $A_c$. In other words, $A_c$ is determined by $c$, and every $b$ where $f(b) = c$ has the exact same final assemby.

%% file: uavgadgets.tex
%\vspace*{-.2cm}
\section{Covert Circuits} \label{sec:uavgadgets}
%\vspace*{-.2cm}

Here we cover the machinery for making covert gadgets and the covert gadgets needed for functional completeness in circuits based on a dual-rail logic implementation: variables, wires, fanouts, and NANDs. We cover a NOT gate as a primitive used in the NAND construction. Traditionally, a crossover is also given, and we discuss why this is unnecessary in Section \ref{sec:uav}. For simplicity, we give some other common gates in Section \ref{sec:fm}.

\begin{figure}[t!]
%\vspace*{-.2cm}
    \centering
	\begin{subfigure}[b]{0.3\textwidth}
	\centering
        \includegraphics[width=1.\textwidth]{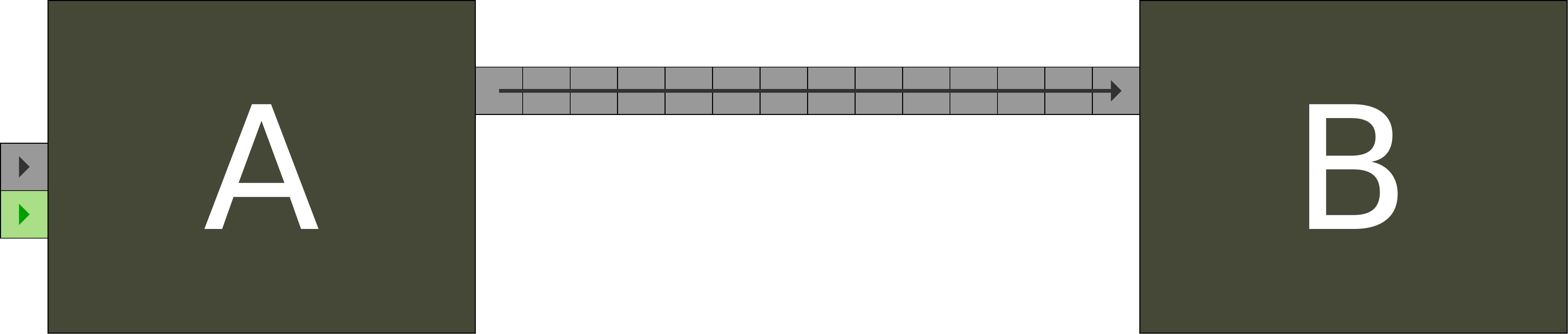}
        \caption{}
        \label{fig:bf1}
    \end{subfigure}
    \begin{subfigure}[b]{0.3\textwidth}
    \centering
        \includegraphics[width=1.\textwidth]{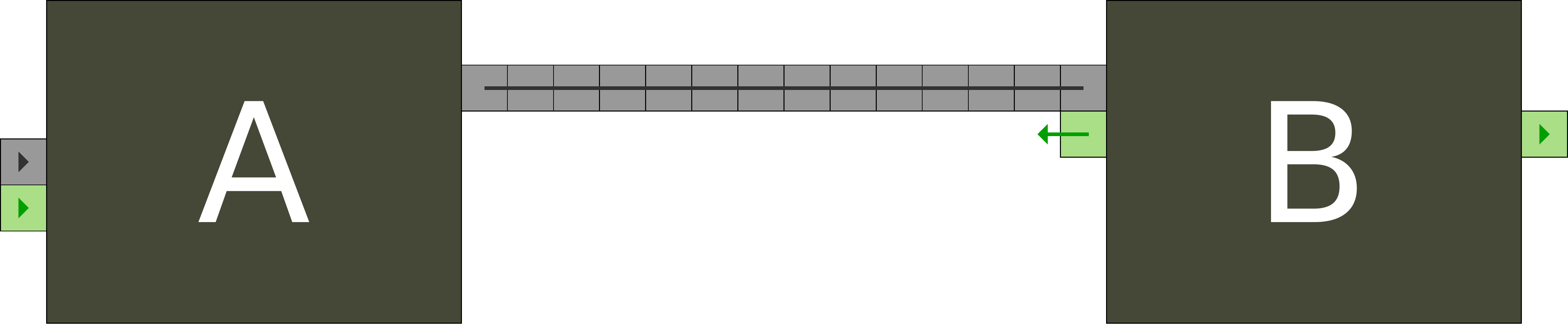}
        \caption{}
        \label{fig:bf2}
    \end{subfigure}
    \begin{subfigure}[b]{0.3\textwidth}
    \centering
        \includegraphics[width=1.\textwidth]{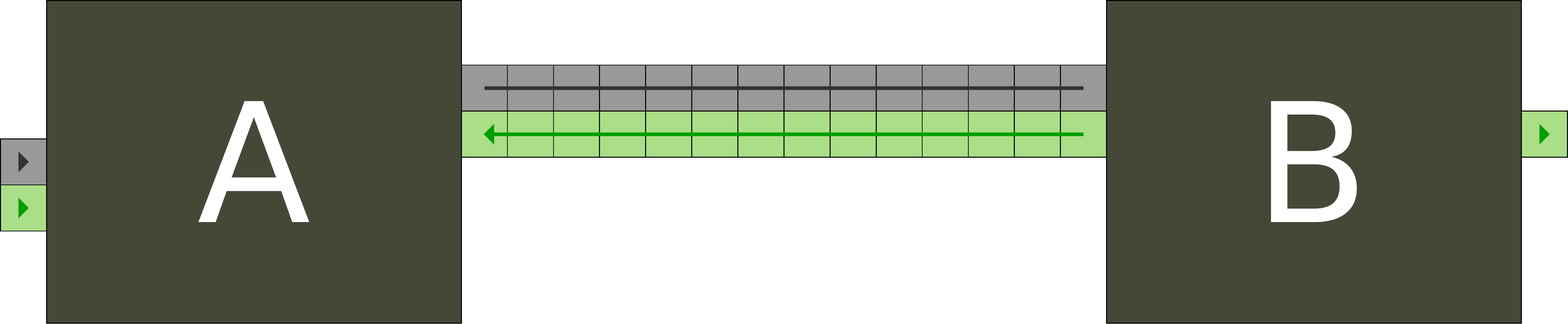}
        \caption{}
        \label{fig:bf3}
    \end{subfigure}
    %\vspace*{-.2cm}
    \caption{An example with two gadgets, A and B, to show how backfilling works in covert computation. (a) If true is output from Gadget A, that wire assembles to the next gadget. (b) Gadget B builds, and based on its function, outputs the true or false wire (false in this case). Once B  received the input (the true wire assembles that then assembles B), it backfills the false wire towards A (the false wire cooperatively assembles the tiles back to the gadget A). (c) The false wire finishes assembling and both Gadget A and B have true and false paths filled. The true output wire of Gadget B will be backfilled from the next gadget. In this way, the input to B/output from A is ``hidden.''} %do not have to be adjacent}
    \label{fig:bf}
    %\vspace*{-.2cm}
\end{figure}

\textbf{Some Conventions.} All solid lines through two neighboring tiles indicate strength-2 glues between them. The arrows indicate the build order (which may branch). Blue single glues are strength 1, and red are strength -1. Following the variable gadget (Figure \ref{fig:ccvar}), all variables have a true and false path adjacent to each other (dual-rail logic), but only one may be traversed at a time until the next gadget. The true value is always to the left or on top of the false value, and for most gadgets, the true input is colored grey while the false input is colored green. Once a variable wire, true or false, reaches the next gadget, the unused variable wire is \emph{backfilled} so that both wires are present. This is a key concept used in all constructions and is further explained in Figure \ref{fig:bf}.

%\vspace*{-.3cm}
\subsection{Variables and Wires}
%\vspace*{-.2cm}

\begin{figure}[t!]
    %\vspace*{-.2cm}
    \centering
    \begin{subfigure}[b]{0.35\textwidth}
	\centering
        \includegraphics[width=.4\textwidth,angle=90]{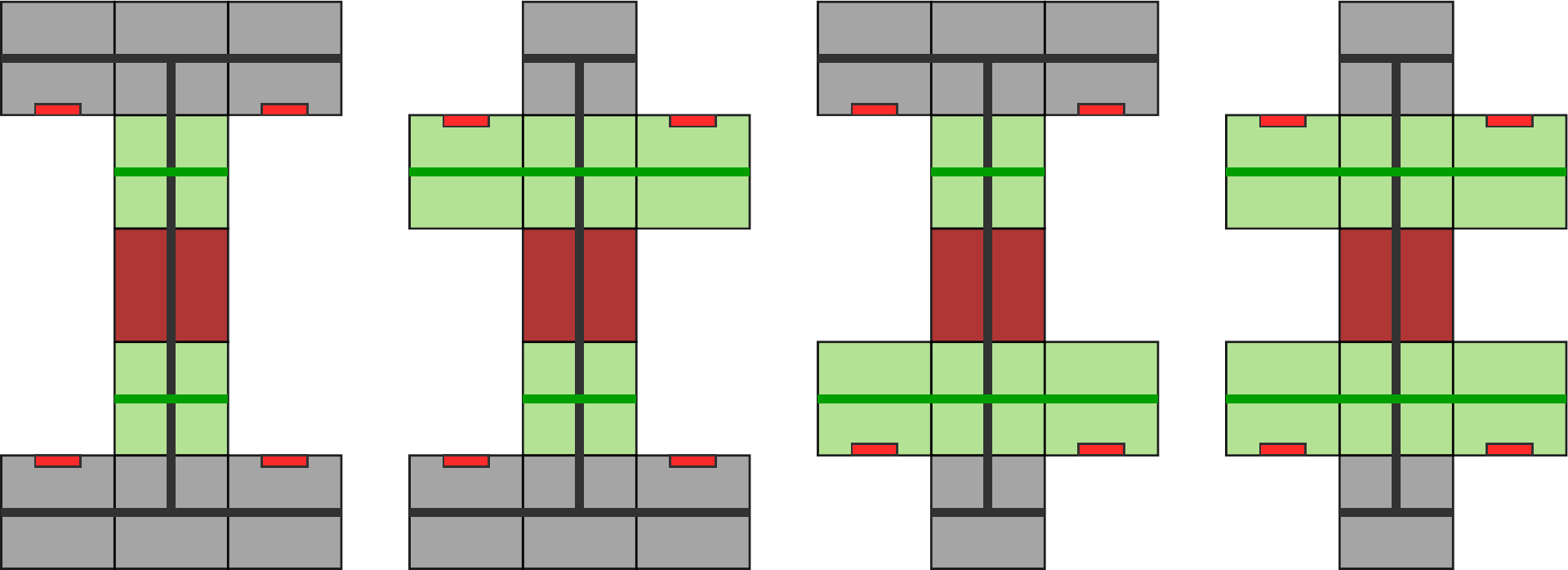}
        \caption{Possible Input Seeds}
        \label{fig:cchaseed1}
    \end{subfigure}
    \hspace*{2.cm}
    \begin{subfigure}[b]{0.35\textwidth}
        \includegraphics[width=1.\textwidth]{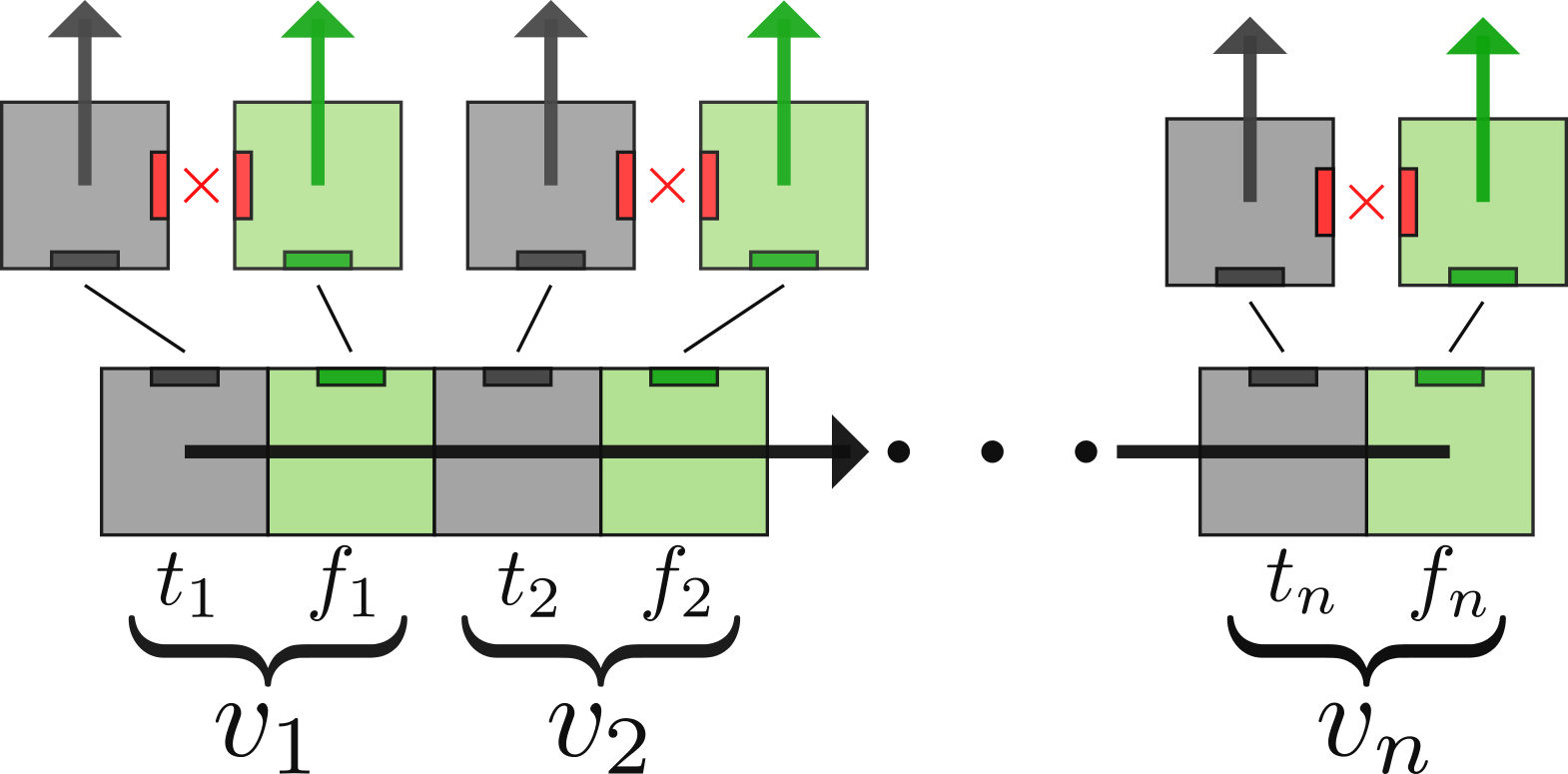}
        \caption{Variable}
        \label{fig:ccvar}
    \end{subfigure}
    %\vspace*{-.2cm}
    \caption{(a) Example of the 4 possible input seeds for a half-adder from Section \ref{subsec:crypt}. (b) Variables are represented by a true and a false line where only one may exist. The variables build off the seed, but only the $t_i$ or the $f_i$ tile may attach due to the negative glue between the two tiles. } %do not have to be adjacent}
    \label{fig:notgadget}
    %\vspace*{-.5cm}
\end{figure}

A variable in our system is represented by two lines of connected tiles where only one exists at a time when the wire is in use (dual rail). Figure \ref{fig:cchaseed1} shows an example of the possible input seeds on 2-bits used in a half-adder. Figure \ref{fig:ccvar} demonstrates how the variables might be set nondeterministically, although generally the specific bits desired would already be attached as part of the input seed (as in Figure \ref{fig:cchaseed1}). Each variable $v_i$ has a sequence of tiles $t_i$ representing a true setting and $f_i$ a false setting. The first tiles have a negative glue of strength $-1$ meaning only the $t_i$ or the $f_i$ tile can attach. The other shown glues are strength 2.
Once the variable is set, the setting travels to the gadget as a \emph{wire}.

The variable setup in Figure~\ref{fig:ccvar} is used in one of two ways:  In the case of providing an input to a covert computation, this variable setup defines the \emph{input template} for the computation, with the seed for a given binary input being the seed assembly with either a true or false tile (but not both) placed at each bit position.  An example system (a half-adder) with a big seed input is shown in %Section \ref{sec:fm}
Figure \ref{fig:ccha}.  Alternatively, the seed begins as a single seed tile that nondeterministically creates a valid input over all possible $n$-bit inputs.  This approach is used in Section~\ref{sec:uav} to show coNP-completeness for unique assembly verification.

%The variable setup in Figure \ref{fig:ccvar} either begins as a seed assembly with either a true or false tile placed for each bit position to provide a valid \emph{input} for a covert computation, or as a single seed tile that nondeterministically creates a valid input over all possible n-bit inputs, allowing us to show coNP-completeness for unique assembly verification in Section~\ref{sec:uav}. In the case of providing an input to a covert computation, To set each of the variables, a \emph{big seed} is used where the desired variable settings are already assembled when mixed into the tiles. This make sense as an idea of a sort of catalyst to begin assembly. We give an example system which uses a big seed in Section \ref{sec:examples} where we show a half-adder.

\begin{figure}[t!]
    %\vspace*{-.2cm}
    \centering
    \begin{subfigure}[b]{0.3\textwidth}
        \includegraphics[height=2.5cm]{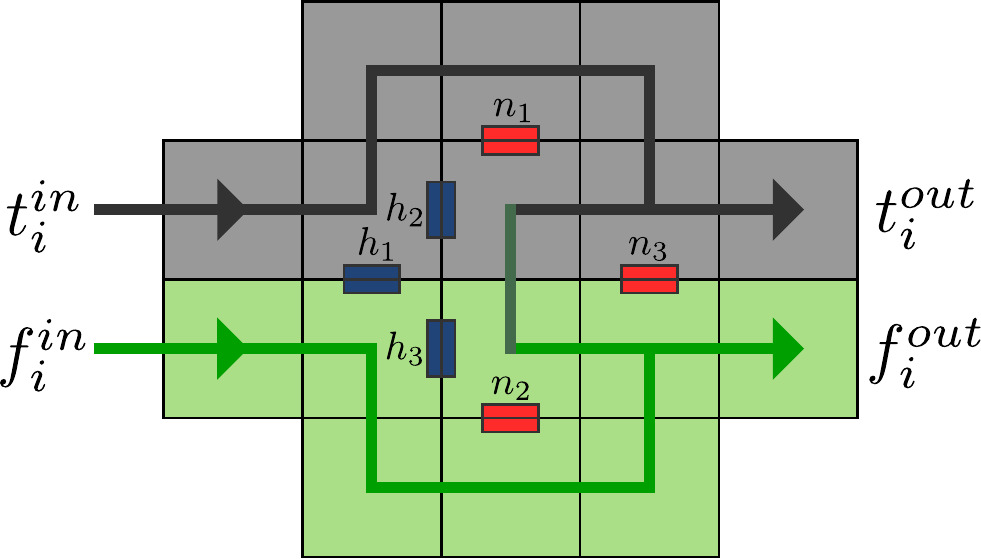}
        \caption{Backfill Stop}
        \label{fig:bfs}
    \end{subfigure}
    \hspace*{2.cm}
	\begin{subfigure}[b]{0.3\textwidth}
        \includegraphics[height=2.5cm]{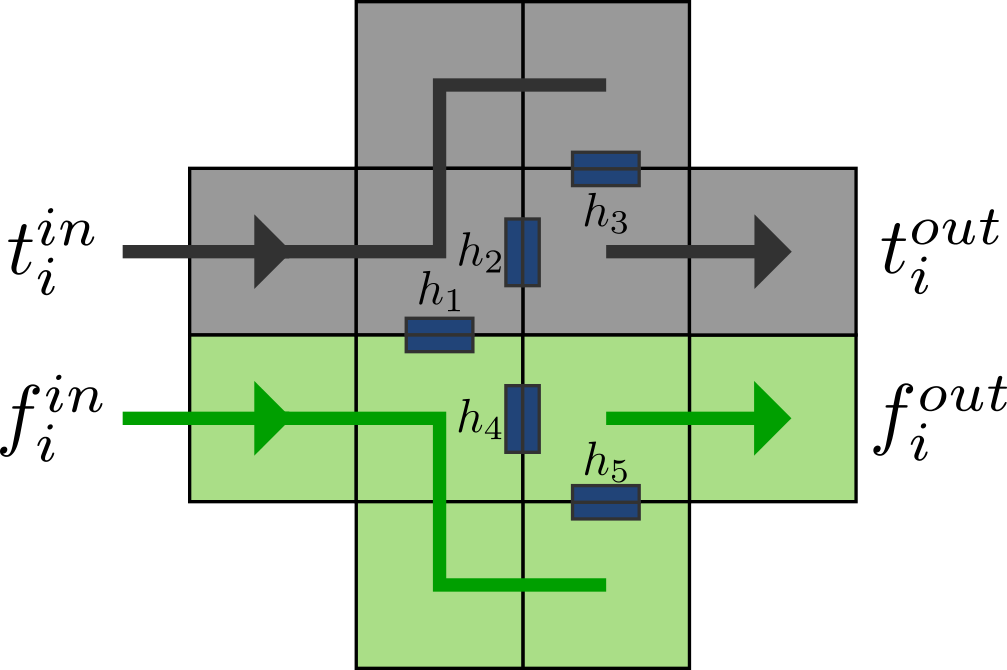}
        \caption{Logic Diode}
        \label{fig:ld}
    \end{subfigure}
    %\vspace*{-.2cm}
    \caption{(a) A backfill stop. This ensures that the wire is allowed to backfill up to a certain point. (b) A gadget referred to as a logic diode. This ensures input from one direction and stops tiles from assembling in the wrong direction. } %do not have to be adjacent}
    \label{fig:bfgadget}
    %\vspace*{-.5cm}
\end{figure}

\subsection{Backfill Stops}
Figure \ref{fig:bfs} shows a \emph{backfill stop}, or simply a \emph{backstop}. This continues a signal, but also backfills the other wire up to this gadget. This ensures that everything up to this gadget has the signals needed to begin backfilling. It has the following properties.

\begin{itemize}
    \item[\textbf{1.}] If $t_i^{in}$ enters, then only $t_i^{out}$ leaves, and $f_i^{in}$ is also populated. Since $t_i^{out}$ attaches the tile with the $n_3$ glue, the tile going to $f_i^{out}$ can not attach. However, the $h_1, h_3$ glues allows a tile from $f_i^{in}$ to attach cooperatively, which allows backfilling of the wire. The $n_2$ glue prevents the wire from attaching another tile towards the output though. 
    
    \item[\textbf{2.}] If $f_i^{in}$ enters, then only $f_i^{out}$ leaves, and $t_i^{in}$ is also populated. Since $f_i^{out}$ attaches the tile with the $n_3$ glue, the tile going to $t_i^{out}$ can not attach. However, the $h_1, h_2$ glues allows a tile from $t_i^{in}$ to attach cooperatively, which allows backfilling of the wire. The $n_1$ glue prevents the wire from attaching another tile towards the output though. 
\end{itemize}

\subsection{Logic Diodes}

Figure \ref{fig:ld} shows what we refer to as a \emph{logic diode}, and prevents timing issues. These appear in every gadget and serve two purposes: if backfilling, this stops the filling at the gadget level so it does not backfill a wire that has not been set, and second it ensures that a gadget must have input from the wire. All shown glues are strength 1 and the lines are strength 2. This gadget is important for later constructions. %The properties of the gadget are in Appendix \ref{subapp:logic}.
It must have these properties.
\begin{itemize}
    \item[\textbf{1.}] If $t_i^{in}$ enters, then only $t_i^{out}$ leaves. This is guaranteed due to $h_2,h_3$ cooperatively attaching the next tile. Without $f_i^{in}$ present, the only tile which could attach is the cooperatively-placed tile from $t_i^{out}$.
    \item[\textbf{2.}] If $f_i^{in}$ enters, then only $f_i^{out}$ leaves. This is guaranteed due to $h_4,h_5$ cooperatively attaching the next tile. Without $t_i^{in}$ present, the only tile which could attach is the cooperatively-placed tile from $f_i^{out}$.
   
    %\item[\textbf{C3.}] If $t_i^{in}$ enters, and the gadget backfills the false path, then the $f_i^{in}$ wire will be backfilled. Given a backfilled false path it will stop at the tile with $h_4,h_5$. However, since the true wire is there, the tile with $h_1,h_4$ can cooperatively attach and backfill the false wire.
    \item[\textbf{3.}] The $t_i^{in}$ wire will be backfilled if and only if $f_i^{in}$ enters. Without $f_i^{in}$ present, a backfilled false path will stop at the tile with $h_2,h_3$. However, with $f_i^{in}$ present, the $h_1,h_2$ tile can cooperatively attach and backfill the true wire.
    %\item[\textbf{C4.}] If $f_i^{in}$ enters, and the gadget backfills the false path, then the $t_i^{in}$ wire will be backfilled. Given a backfilled true path it will stop at the tile with $h_2,h_3$. However, since the false wire is there, the tile with $h_1,h_2$ can cooperatively attach and backfill the true wire.
     \item[\textbf{4.}] The $f_i^{in}$ wire will be backfilled if and only if $t_i^{in}$ enters. Without $t_i^{in}$ present, a backfilled false path will stop at the tile with $h_4,h_5$. With $t_i^{in}$ present, the tile with $h_1,h_4$ can cooperatively attach and backfill the false wire.
\end{itemize}

%\vspace*{-.3cm}
\subsection{Covert NOT Gadget}
%\vspace*{-.2cm}

\begin{figure}[t!]
    %\vspace*{-.2cm}
    \centering
    \begin{subfigure}[b]{0.33\textwidth}
        \includegraphics[width=1.\textwidth]{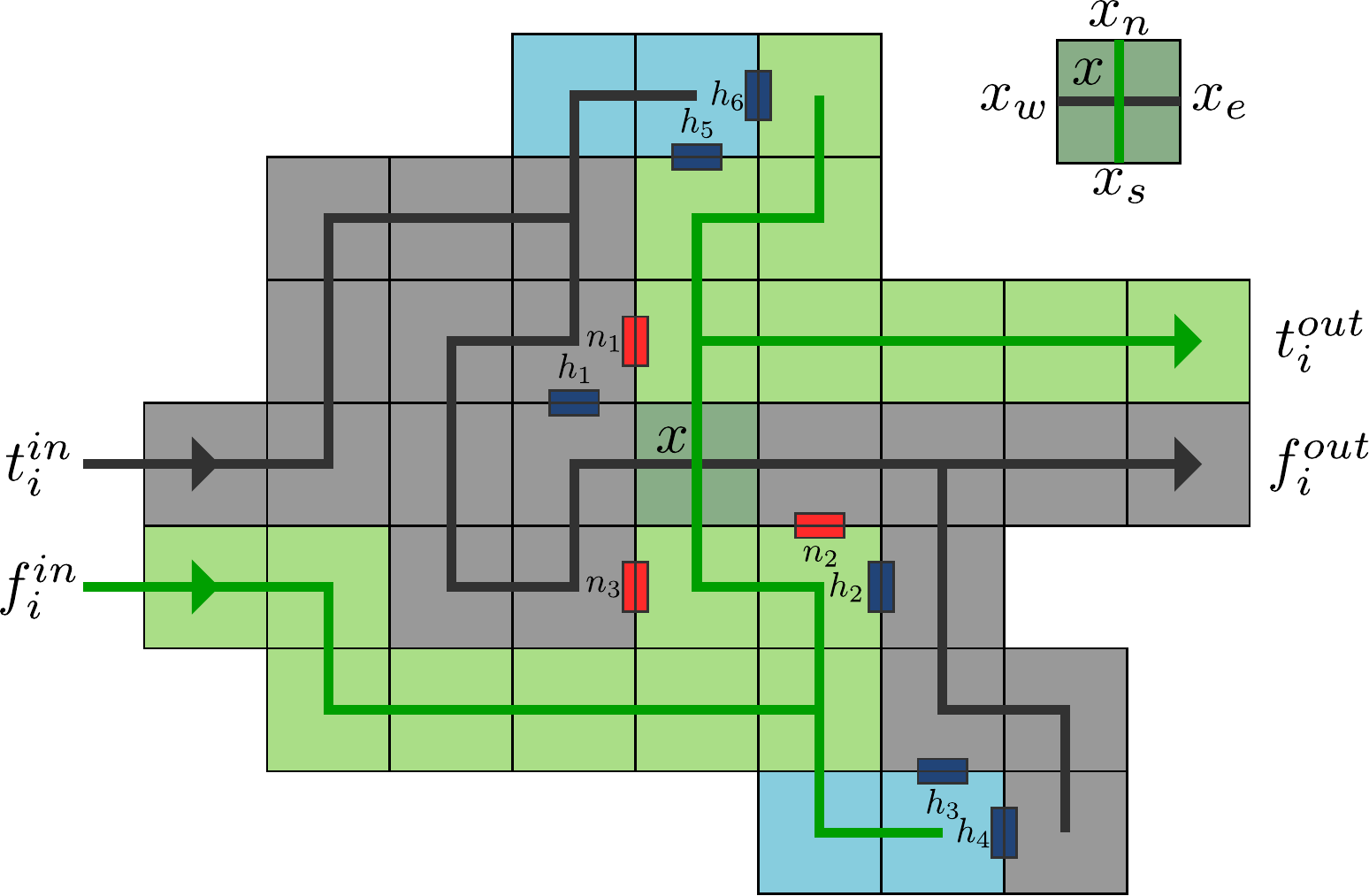}
        \caption{Basic NOT}
        \label{fig:notb}
    \end{subfigure}
    %\hspace*{.2cm}
    \begin{subfigure}[b]{0.33\textwidth}
        \includegraphics[width=1.\textwidth]{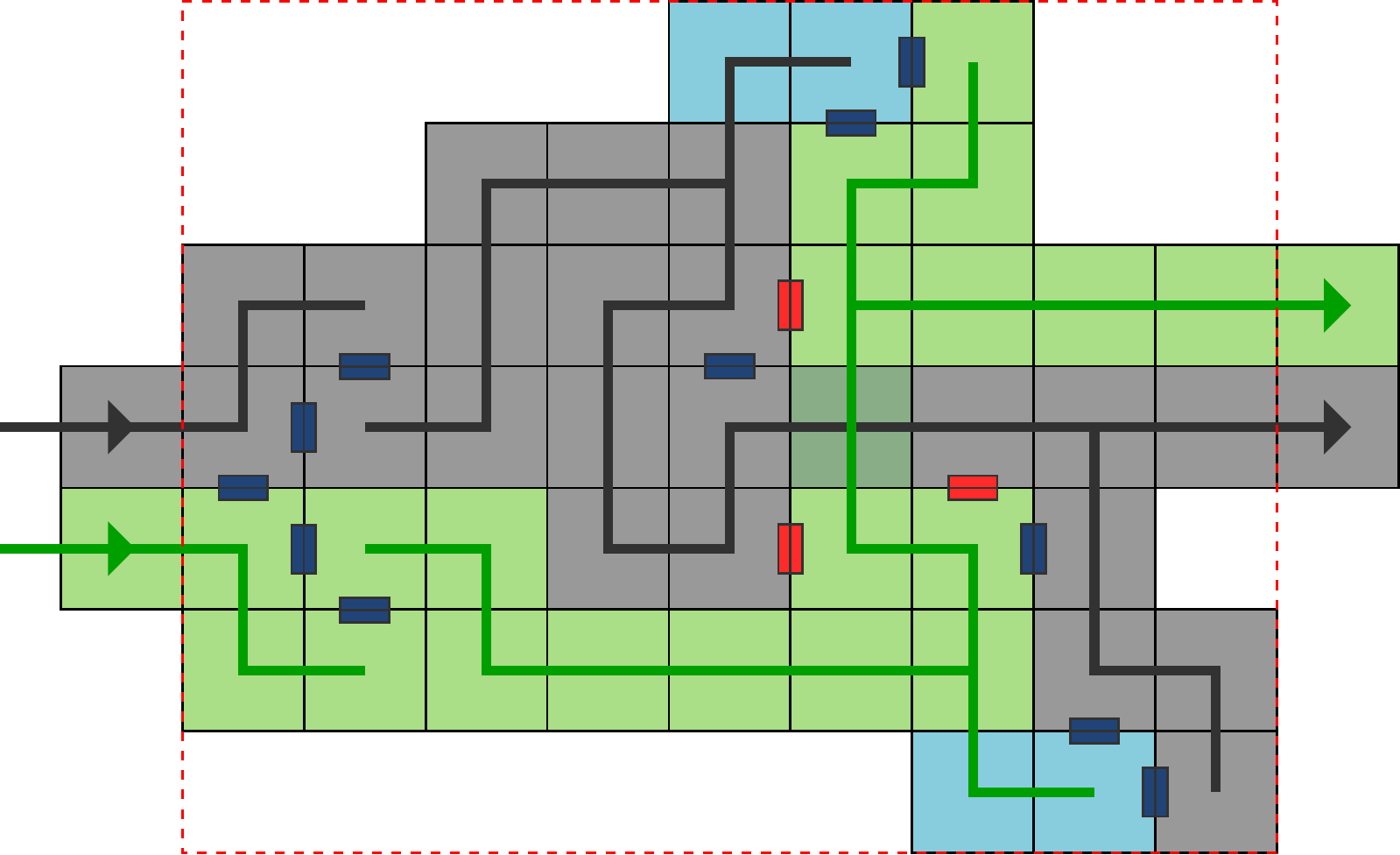}
        \caption{NOT}
        \label{fig:ccnot}
    \end{subfigure}
    %\hspace*{.2cm}
	\begin{subfigure}[b]{0.32\textwidth}
        \includegraphics[width=1.\textwidth]{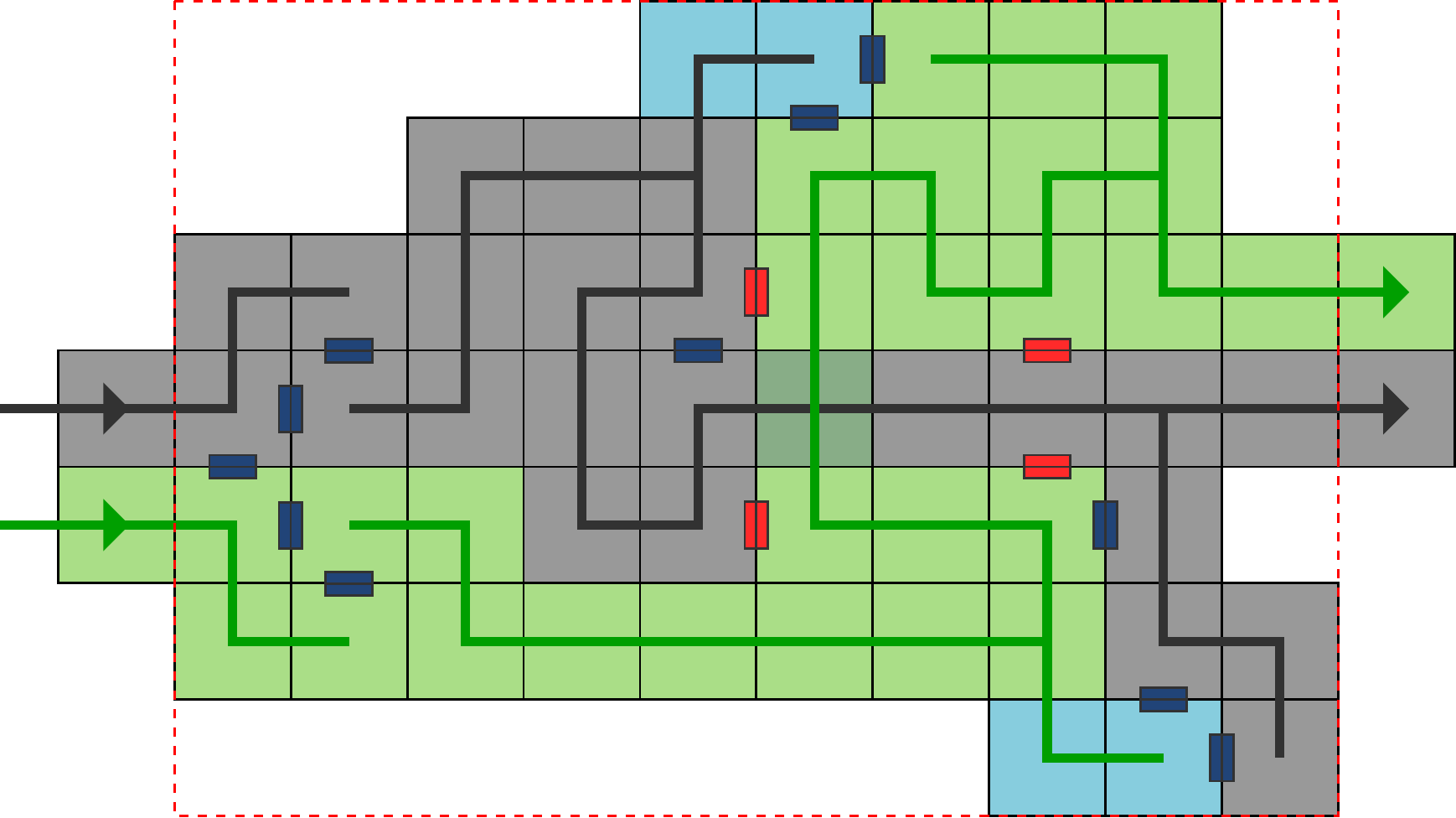}
        \caption{H-NOT}
        \label{fig:ccnoth}
    \end{subfigure}
    %\vspace*{-.2cm}
    \caption{
    (a) Basic NOT gate (b) NOT gate with the logic diode on the input (c) A covert NOT gate with an additional negative horizontal glue on the output to prevent incorrect backfilling. This modification is needed when using this gate for the construction of the NAND gate.
    %(b) A covert NOT gate with an additional negative vertical glue on the output to prevent incorrect backfilling. Needed for the NAND gate.
    }
    \label{fig:notalt}
    %\vspace*{-.6cm}
\end{figure}

The first covert gadget we introduce is a NOT gadget.  This gadget displays some of the key insights needed for covert computation, such as how blocking with negative glue adds power to the system.  The NOT gadget is also used as a submodule within our NAND gadget.  The NOT gadget is shown in Figure \ref{fig:notb}, and Figure \ref{fig:ccnot} is the NOT gadget with the logic diode added on the input to ensure no backfill happens unless the gadget has receieved input. In the NAND gadget, we also use the same structure as the NOT gadget, but with an additional negative glue. This is shown in Figure \ref{fig:ccnoth} and will be discussed when needed.

\begin{figure}[t!]
    %\vspace*{-.2cm}
    \centering
    \begin{subfigure}[b]{0.24\textwidth}
        \includegraphics[width=1.\textwidth]{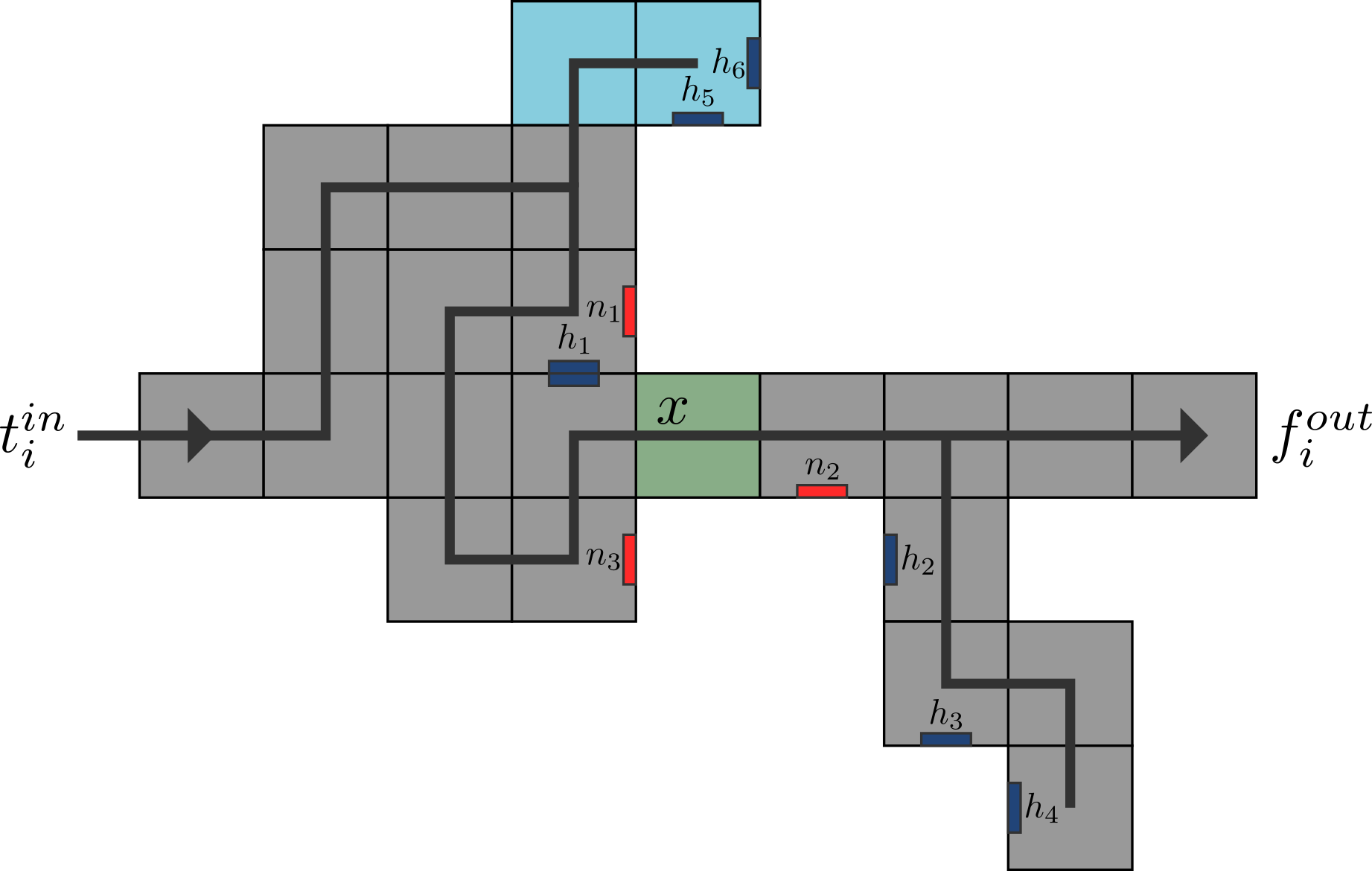}
        \caption{True Input 1}
        \label{fig:ccnott1}
    \end{subfigure}
    %5\hspace*{.2cm}
	\begin{subfigure}[b]{0.24\textwidth}
        \includegraphics[width=1.\textwidth]{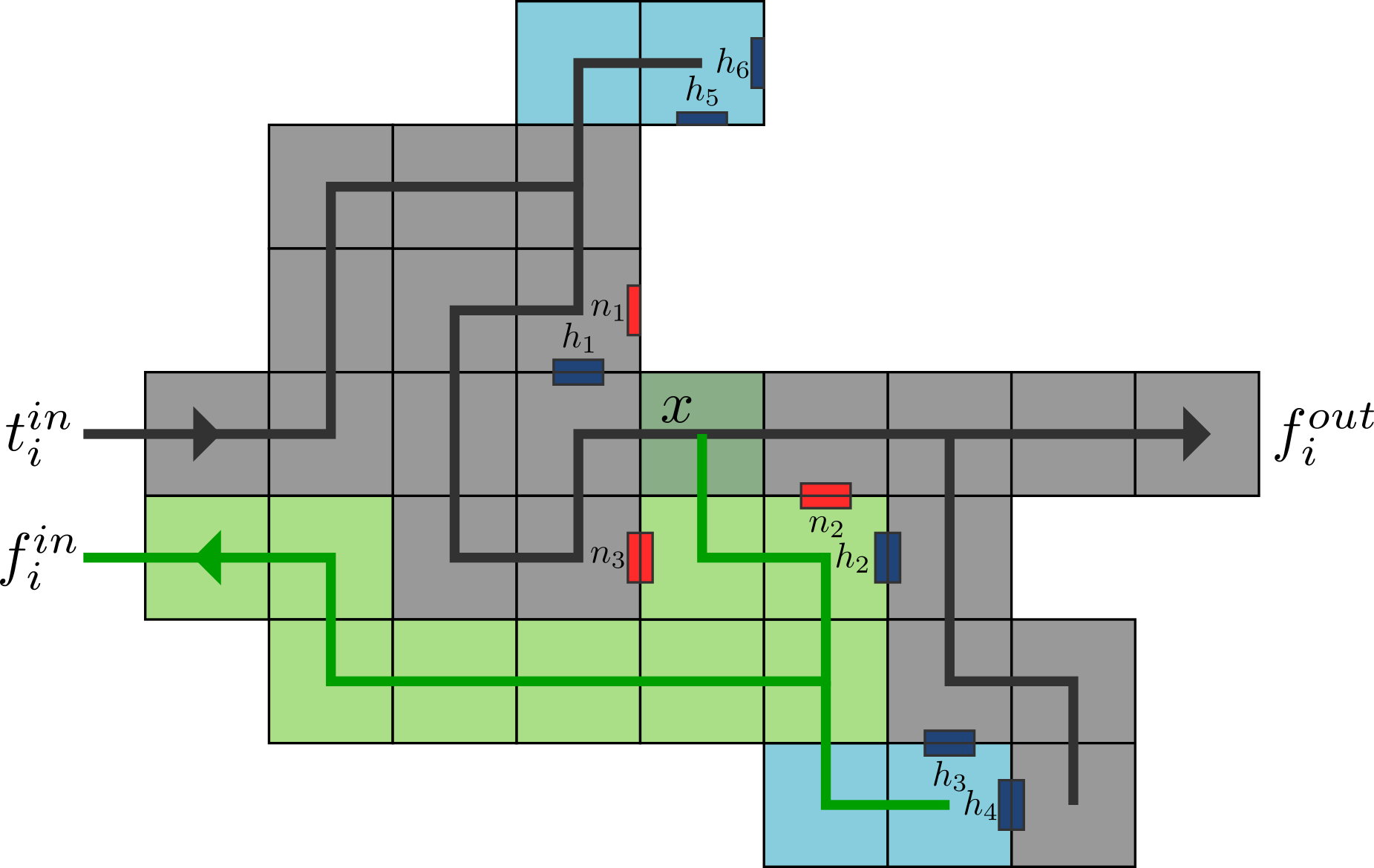}
        \caption{True Input 2}
        \label{fig:ccnott2}
    \end{subfigure}
      %  \hspace*{.1cm}
	\begin{subfigure}[b]{0.24\textwidth}
        \includegraphics[width=1.\textwidth]{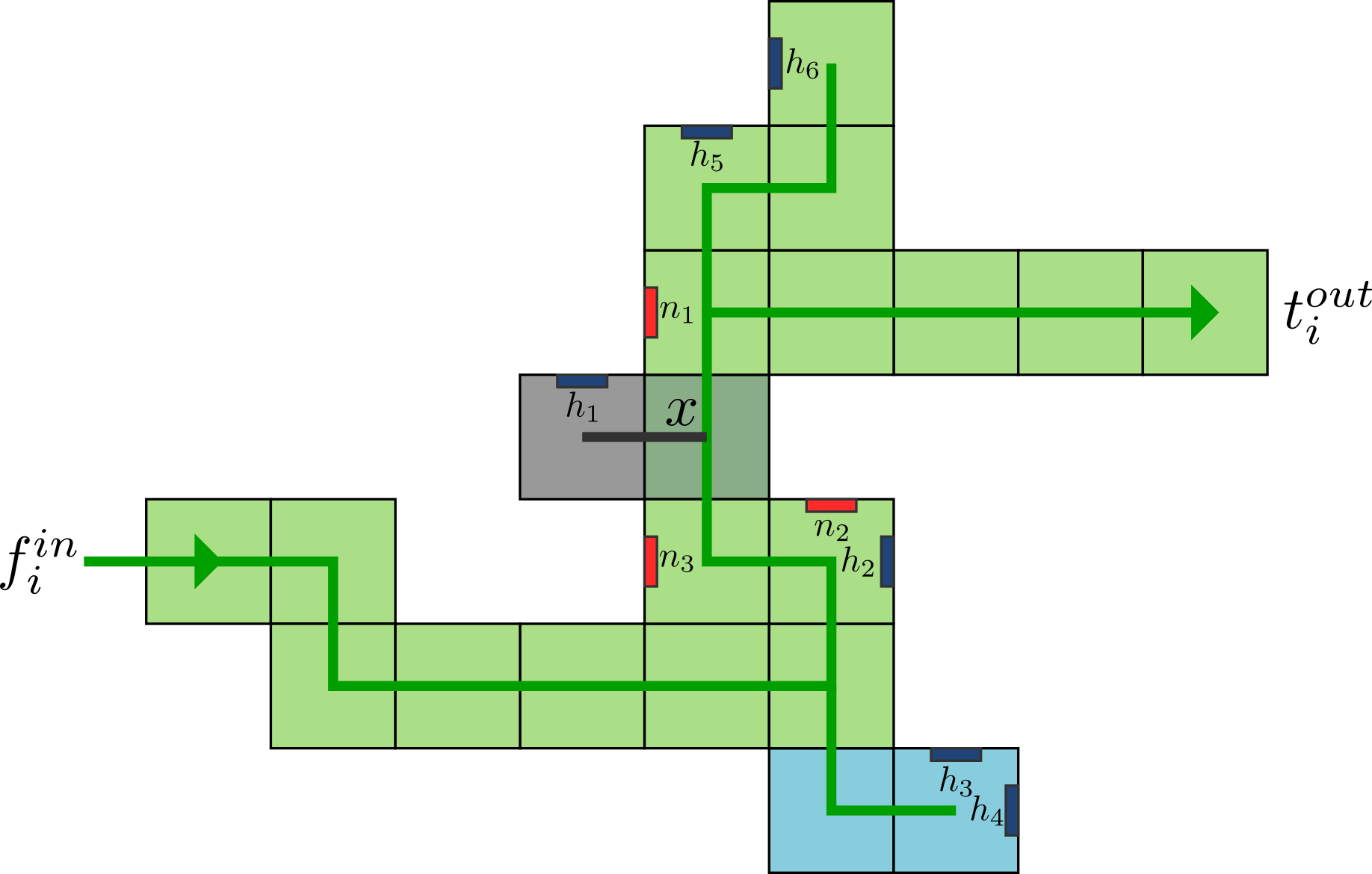}
        \caption{False Input 1}
        \label{fig:ccnotf1}
    \end{subfigure}
    %\hspace*{.2cm}
	\begin{subfigure}[b]{0.24\textwidth}
        \includegraphics[width=1.\textwidth]{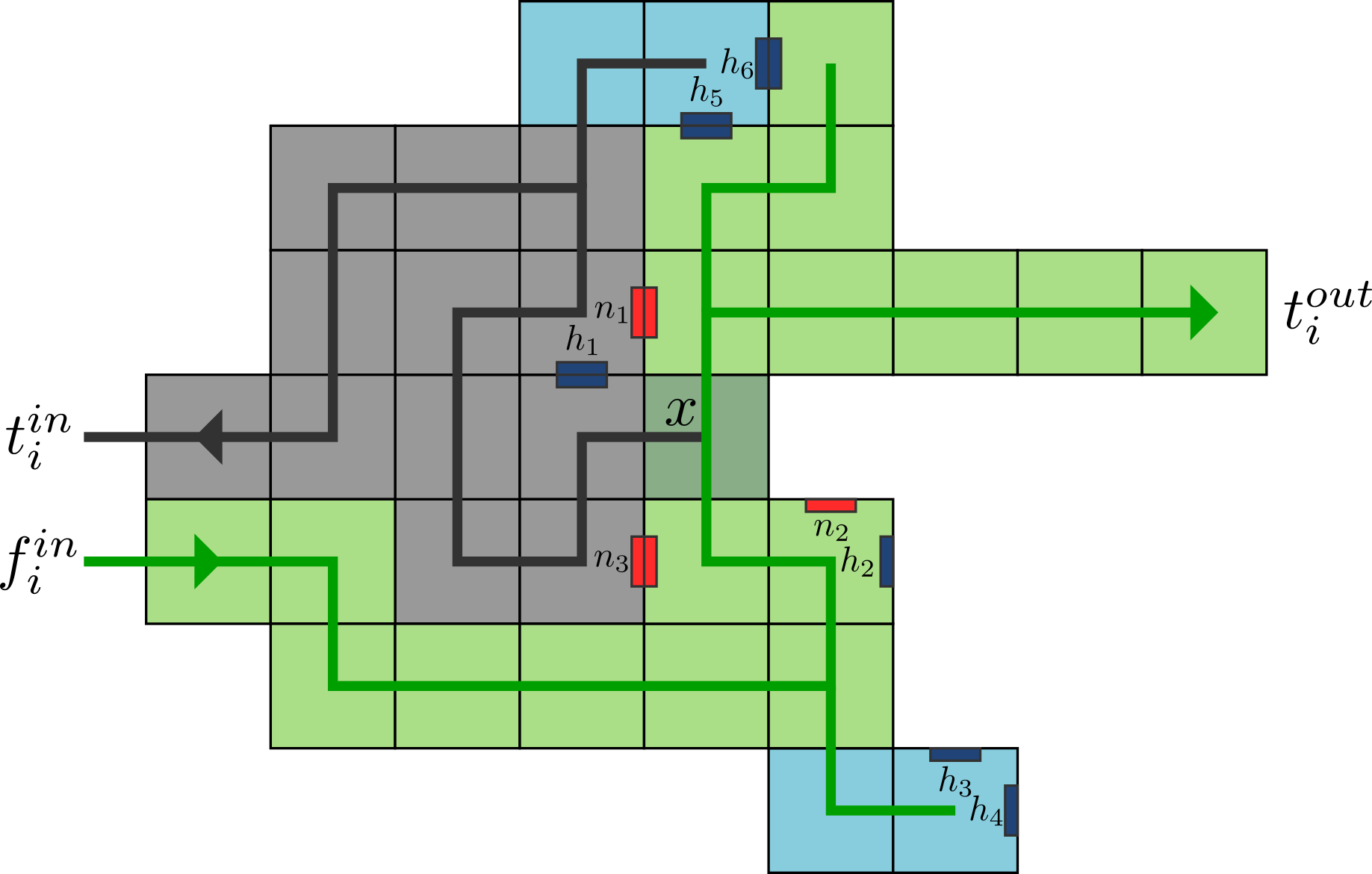}
        \caption{False Input 2}
        \label{fig:ccnotf2}
    \end{subfigure}
    %\vspace*{-.2cm}
    \caption{
    (a) A NOT gadget with true input $t_i^{in}$ and output $f_i^{out}$. The true output can not place from tile $x$ due to the negative glues $n_1$ and $n_3$ of strength $-1$. (b) Once the NOT gadget passes the false output, glues $h_3,h_4$ cooperatively allow the false portion and wire to backfill. Glue $h_2$ is needed to fill in the tile with $n_2$. (c) A NOT gadget with false input $f_i^{in}$ and output $t_i^{out}$. The false output can not attach due to the negative glue $n_2$. The tile to the west of $x$ may attach, but due to glue $n_3$, no other tile can attach. (d) Once the NOT gadget passes the true output, glues $h_5,h_6$ allow the true portion and wire to backfill. Glue $h_1$ is needed to counteract the $n_1$ glue when backfilling that tile.
    }
    \label{fig:notprf}
    %\vspace*{-.2cm}
\end{figure}

% \begin{lemma}
% The covert NOT works.
% \end{lemma}

%\begin{proof}
Given the variables and wires work as shown, the difficulty in a dual-rail NOT is that there must be at least one crossing tile that both the true and false paths place. This tile can be thought of as where the signals cross or switch.
Figure \ref{fig:notb} shows the basic NOT gadgets, and the tile shared by both paths is labelled $x$. The negative glues allow blocking around this tile so that only one path is possible once $x$ is placed.

%The specific properties that must hold are covered in Appendix \ref{subapp:NOT}.
The properties of the NOT gadget guarantee that it works correctly and that the gadget is covert (the gadget looks indistinguishable before the output regardless of the input), and that the backfill works correctly. Figure \ref{fig:notprf} discusses these elements and walks through how the true/false inputs block and crossover correctly. The figure does not show the logic diode though.

In verifying that the gadget works as intended, we must verify six properties. The first two conditions guarantee that the gadget works correctly. The second two conditions are needed along with the last two to guarantee the covertness of the gadget, i.e., the gadget looks indistinguishable before the output regardless of the input. The final two conditions also verify that the backfill from future gadgets will work correctly, and no trace of the build path will be evident.

\begin{itemize}
    \item[\textbf{1.}] If $t_i^{in}$ enters a NOT gadget, it results in $f_i^{out}$ and not $t_i^{out}$.
    Figure \ref{fig:ccnott1} shows the gadget in this case with true input $t_i^{in}$ and output $f_i^{out}$. The true output can not place from tile $x$ due to the negative glues $n_1$ and $n_3$ of strength $-1$. Given the build order, we are guaranteed $f_i^{out}$ and that $t_i^{out}$ can not build.

    \item[\textbf{2.}] If $f_i^{in}$ enters a NOT gadget, it results in $t_i^{out}$ and not $f_i^{out}$.
    Figure \ref{fig:ccnotf1} shows the gadget in this case$-$ with false input $f_i^{in}$ and output $t_i^{out}$. The false output can not attach due to the negative glue $n_2$. The tile to the west of $x$ may attach, but due to glue $n_3$, no other tile can attach. Given the build order, we are guaranteed $t_i^{out}$ and that $f_i^{out}$ can not build.

    \item[\textbf{3.}] The $t_i^{in}$ wire will be backfilled up to tile $x$ if and only if $f_i^{in}$ enters a NOT gadget and $t_i^{out}$ leaves. Figure \ref{fig:ccnotf2} shows the desired result. Glues $h_5,h_6$ allow the true portion and wire to backfill. Glue $h_1$ is needed to counteract the $n_1$ glue when backfilling that tile. Then the logic diode ensures $f_i^{in}$ is present to backfill the wire.
    
    \item[\textbf{4.}] The $f_i^{in}$ wire will be backfilled up to tile $x$ if and only if $t_i^{in}$ enters a NOT gadget and $f_i^{out}$ leaves. Figure \ref{fig:ccnott2} shows the desired result. Glues $h_3,h_4$ allow the true portion and wire to backfill.  Glue $h_2$ is needed to fill in the tile with $n_2$. Then the logic diode ensures $t_i^{in}$ is present to backfill the wire.
    
    %\item[\textbf{4.}] If $t_i^{in}$ enters a NOT gadget and $f_i^{out}$ leaves, the gadget and $f_i^{in}$ is backfilled up to tile $x$. Figure \ref{fig:ccnott2} shows the desired result. Glues $h_3,h_4$ cooperatively allow the false portion and wire to backfill. Glue $h_2$ is needed to fill in the tile with $n_2$.

    \item[\textbf{5.}] If the gadget resulted in $f_i^{out}$, a future gadget can backfill $t_i^{out}$ and the gadget will be complete. If the gadget is in the configuration of Figure \ref{fig:ccnott2}, the true wire can directly backfill until the tile directly above tile $x$. The glue $n_1$ would prevent this tile from placing except $x$ will be there and the tile can cooperatively attach using the north glue of $x$ and the south glue of the backfilling wire.

    \item[\textbf{6.}] If the gadget resulted in $t_i^{out}$, a future gadget can backfill $f_i^{out}$ and the gadget will be complete. If the gadget is in the configuration of Figure \ref{fig:ccnotf2}, the false wire can directly backfill until the tile directly east of tile $x$. The glue $n_2$ would prevent this tile from placing except $x$ is there and the tile can cooperatively attach using the east glue of $x$ and the west glue of the backfilling wire.
\end{itemize}

%\paragraph{Case 1.} Figure \ref{fig:ccnott1} shows the gadget in this case- with true input $t_i^{in}$ and output $t_i^{out}$. The false output can not place from tile $x$ due to the negative glues $n_1$ and $n_3$ of strength $-1$. Given the build order, we are guaranteed $t_i^{out}$ and that $f_i^{out}$ can not build.

%\paragraph{Case 2.} Figure \ref{fig:ccnotf1} shows the gadget in this case- with false input $f_i^{in}$ and output $f_i^{out}$. The true output can not attach due to the negative glue $n_2$. The tile to the west of $x$ may attach, but due to glue $n_3$, no other tile can attach. Given the build order, we are guaranteed $f_i^{out}$ and that $t_i^{out}$ can not build.

%\paragraph{Case 3.}

%\paragraph{Case 4.}
%\end{proof}

\begin{figure}[t!]
    %\vspace*{-.2cm}
    \centering
	\begin{subfigure}[b]{1.\textwidth}
        \includegraphics[width=1.\textwidth]{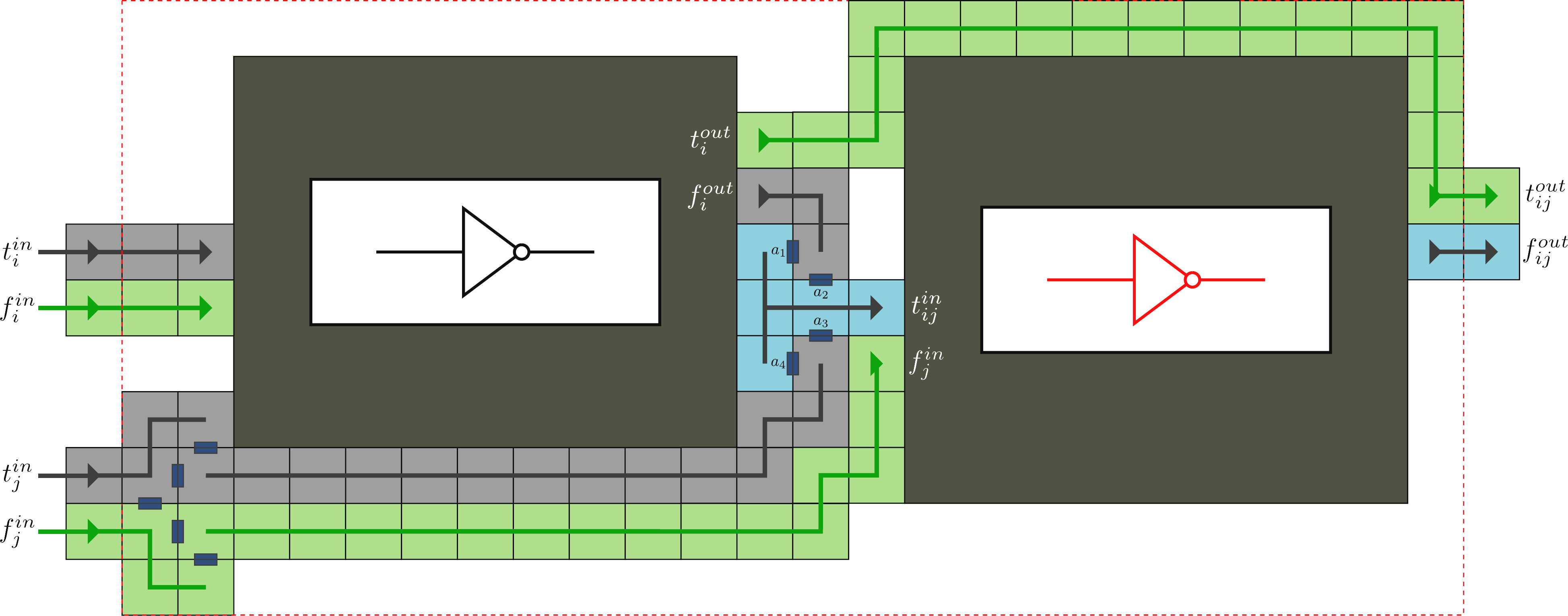}
        \caption{NAND Block Diagram}
        \label{fig:ccnandb}
    \end{subfigure}
      %  \hspace*{.3cm}
      
	\begin{subfigure}[b]{1.\textwidth}
        \includegraphics[width=1.\textwidth]{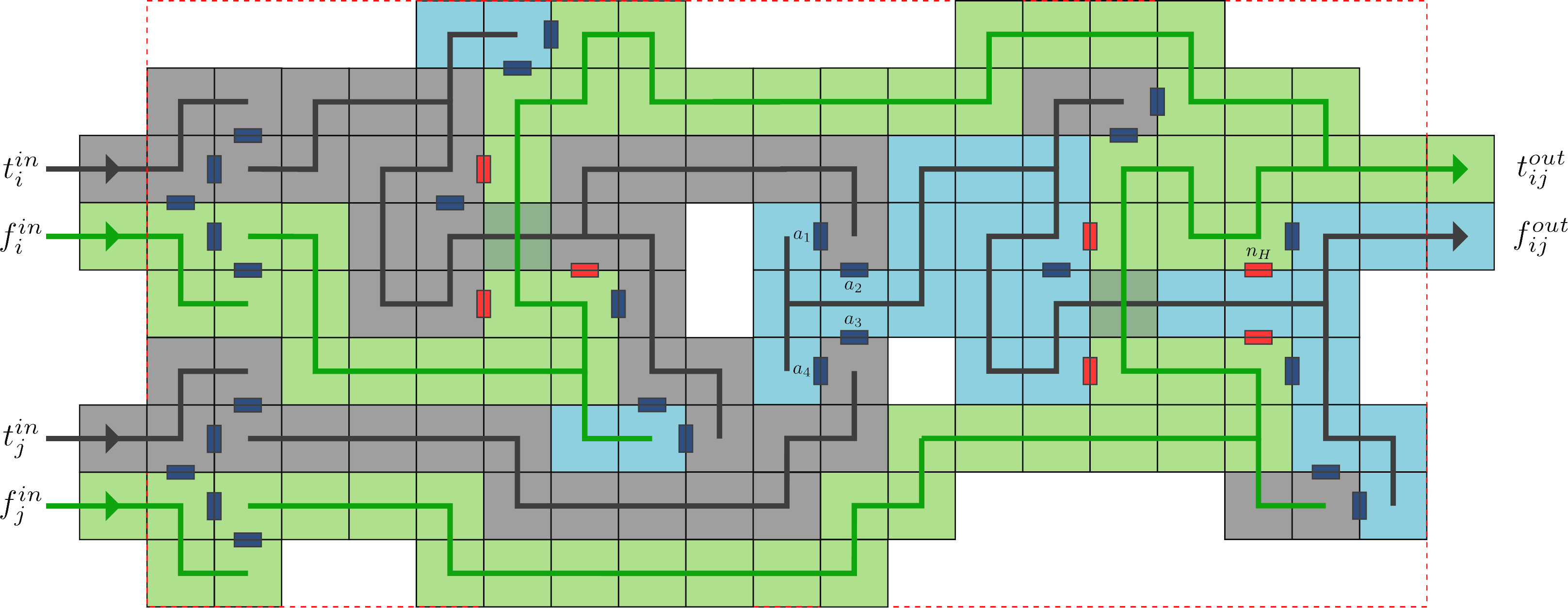}
        \caption{NAND}
        \label{fig:ccnand}
    \end{subfigure}
    %\vspace*{-.2cm}
    \caption{(a) Diagram of the covert NAND gate with NOTs shown as blocks. The boxes for the NOT blocks are shown outlined in Figures \ref{fig:ccnot} and \ref{fig:ccnoth}. The left box is the standard NOT gadget and the right box is the H-NOT gadget. (b) The full NAND gate with the two NOT gadgets filled in and compacted.}
    \label{fig:nandgadget}
    %\vspace*{-.2cm}
\end{figure}

%\vspace*{-.3cm}
\subsection{Covert NAND Gadget}
%\vspace*{-.2cm}

The basic idea for the NAND gadget is to compare if both inputs are true, but because of the planarity constraints, we need to ``flip'' one of the inputs using a covert NOT before comparing. 
%to flip one of the inputs using a covert NOT, and then we can compare the two true input lines to see if both inputs are true. 
Since a NAND is false only when both inputs are true, this is the only path that should result in a false output. The basic idea for the gadget is shown in Figure \ref{fig:ccnandb} with a representative block for the NOT gadget already discussed. The second NOT block is the modified NOT gadget (H-NOT) from Figure \ref{fig:ccnoth}. Both false inputs are routed to the true output. One must go through another NOT in order to flip to the top output position, while the other false line skips this NOT and ties directly to the true output.
%There are several things that could assemble incorrectly and are handled in various ways.
%For clarity, we discuss the block diagram first, then address a few technical issues in the full version.
%The full list of properties necessary and sufficient for the covert NAND is given in Appendix \ref{subapp:NAND}.
Once we flip the top input, we can use cooperative binding to compare the two true inputs, and only if both are true do we send it as true into the second NOT block (so the gadget outputs false). All other input combinations output true.

We will show why NOT and H-NOT are both necessary. Looking at Figure \ref{fig:ccnand}, the negative glue $n_H$ is necessary in H-NOT to ensure that $t_i^{out}$, which skips the second NOT gadget, does not set the output $t_{ij}^{out}$, and then also set $f_{ij}^{out}$ based on the assembly order. Essentially, this protects from incorrect backfilling and setting both outputs. However, the $n_H$ glue should not exist in the standard NOT gadget, or it may backfill and could cause a tile to break off depending on build order. Given we are using the growth-only aTAM, this would not be allowed. It is possible to create a single NOT that incorporates these properties, but we prefer to avoid the added complexity.

Finally, the logic diodes on the inputs (Figure \ref{fig:ld}) ensure that if we only have one input, the gadget does not backfill down the other input wire. Even if the gadget has already been set, that input will wait until either the true or false wire comes before backfilling the wire.

Given the complexity of the NAND gadget, there are more issues related to its function that must be considered compared to the previous gadgets. The properties that it must have are as follows.
\begin{itemize}
    \item[\textbf{1.}] If wires $t_i^{in}$ and $t_j^{in}$ are set, then the wire $f_{ij}^{out}$ should leave the gadget. Wire $t_i^{in}$ exits the first NOT gadget as $f_i^{out}$. This wire, $f_i^{out}$, and $t_j^{in}$ both stop and expose glues $a_2$ and $a_3$, respectively. Both are strength 1 glues, and thus the tile with glues $a_2$ and $a_3$ can only attach if both the glues are exposed. Thus, only if wires $t_i^{in}$ and $t_j^{in}$ are set, will wire $t_{ij}^{in}$ ever enter the H-NOT gadget, which results in the wire $f_{ij}^{out}$ as the gadget output.

    \item[\textbf{2.}] Given $f_i^{in}$ or $f_j^{in}$, the wire $t_{ij}^{out}$ leaves the gadget. If wire $f_i^{in} = t_i^{out}$ is set, this is tied directly to the output wire $t_{ij}^{out}$. If the wire $f_j^{in}$ is set, it comes in as the false input of the second H-NOT gadget, which means it leaves as $t_{ij}^{out}$ by the validity of the NOT gadget.

    \item[\textbf{3.}] The wires $t_i^{in}$ and $f_j^{in}$ ($f_i^{in}$ and $t_j^{in}$ input) are backfilled if $t_{ij}^{out}$ left the gadget. If $t_{ij}^{out}$ left the gadget, then it directly can backfill both the $t_i^{out}$ wire of the NOT gadget and the $f_j^{in}$ wire if not already filled. Thus, $f_j^{in}$ is filled. Since the input line $t_i^{out}$ of the NOT is filled, it backfills $t_i^{in}$ as shown for the NOT gadget. Both wires are then backfilled through cooperative attachment with the presence of the other input in the logic diode. 
    
    \item[\textbf{4.}] The wires $f_i^{in}$ and $t_j^{in}$ ($t_i^{in}$ and $f_j^{in}$ input) are backfilled if $t_{ij}^{out}$ left the gadget. If $t_{ij}^{out}$ left the gadget, then it directly can backfill both the $t_i^{out}$ wire of the NOT gadget and the $f_j^{in}$ wire if not already filled. The true output of the H-NOT gadget backfills and glues $a_3, a_4$ allow the cooperative attachment that backfills $t_j^{in}$. The NOT gadget with the true input would backfill $f_i^{in}$. Both wires are then backfilled through cooperative attachment with the presence of the other input in the logic diode.  
    
    \item[\textbf{5.}] The wires $t_i^{in}$ and $t_j^{in}$ ($f_i^{in}$ and $f_j^{in}$ input) are backfilled if $t_{ij}^{out}$ left the gadget. If $t_{ij}^{out}$ left the gadget, then it directly can backfill both the $t_i^{out}$ wire of the NOT gadget and the $f_j^{in}$ wire if not already filled. The NOT gadget with $f_i^{in}$ input will backfill $t_i^{in}$. The true output of the H-NOT gadget backfills and glues $a_3, a_4$ allow the cooperative attachment that backfills $t_j^{in}$. Both wires are then backfilled through cooperative attachment with the presence of the other input in the logic diode.

    \item[\textbf{6.}] The wires $f_i^{in}$ and $f_j^{in}$ ($t_i^{in}$ and $t_j^{in}$ input) are backfilled if $f_{ij}^{out}$ left the gadget. The NOT gadget with the true input would backfill $f_i^{in}$. The H-NOT gadget will backfill the false input line, which is $f_j^{in}$. Both wires are then backfilled through cooperative attachment with the presence of the other input in the logic diode.  
    
    \item[\textbf{7.}] The wire $f_i^{out}$ is always placed. The glues $a_1,a_2$ ensure that if all previous properties hold, the false output wire ($f_i^{out}$) is also filled. Stepping through properties 3-6, it is possible for the gadget to backfill the input wires correctly without always placing these tiles.
    
    \item[\textbf{8.}] The growth-only constraint is not violated with the negative glues. This can only happen given a stable assembly where a tile attaches with a negative glue that destabilizes part of the assembly. The additional negative glue $n_H$ could do this if the green tile is placed after the blue tile, however, the build path is intentional to ensure this can not happen. If the wire $f_{ij}^{out}$ were placed and $t_{ij}^{out}$ is backfilled, the tile with $n_H$ would be the last tile that could attach and the assembly would never be unstable.

    \item[\textbf{9.}] The gadget does not behave incorrectly with only one input. The logic diode guarantees the backfilling never goes beyond the gadget. If one input is false, the NAND can send the $t_{ij}^{out}$ wire and backfill the NAND gadget without having yet receieved the second input. When the second wire does arrives, that wire is backfilled. With one true input, the gadget will just be waiting at the tile needing both $a_2$ or $a_3$. $t_i^{in}$ will also backfill $f_i^{in}$ even with only that input. Given $t_j^{in}$, the false line is not backfilled until both inputs are there.
    
\end{itemize}

%\vspace*{-.3cm}
\subsection{Covert FANOUT Gadget}
%\vspace*{-.2cm}

\begin{figure}[t!]
    %\vspace*{-.2cm}
    \centering
	\begin{subfigure}[b]{0.6\textwidth}
        \includegraphics[width=1.\textwidth]{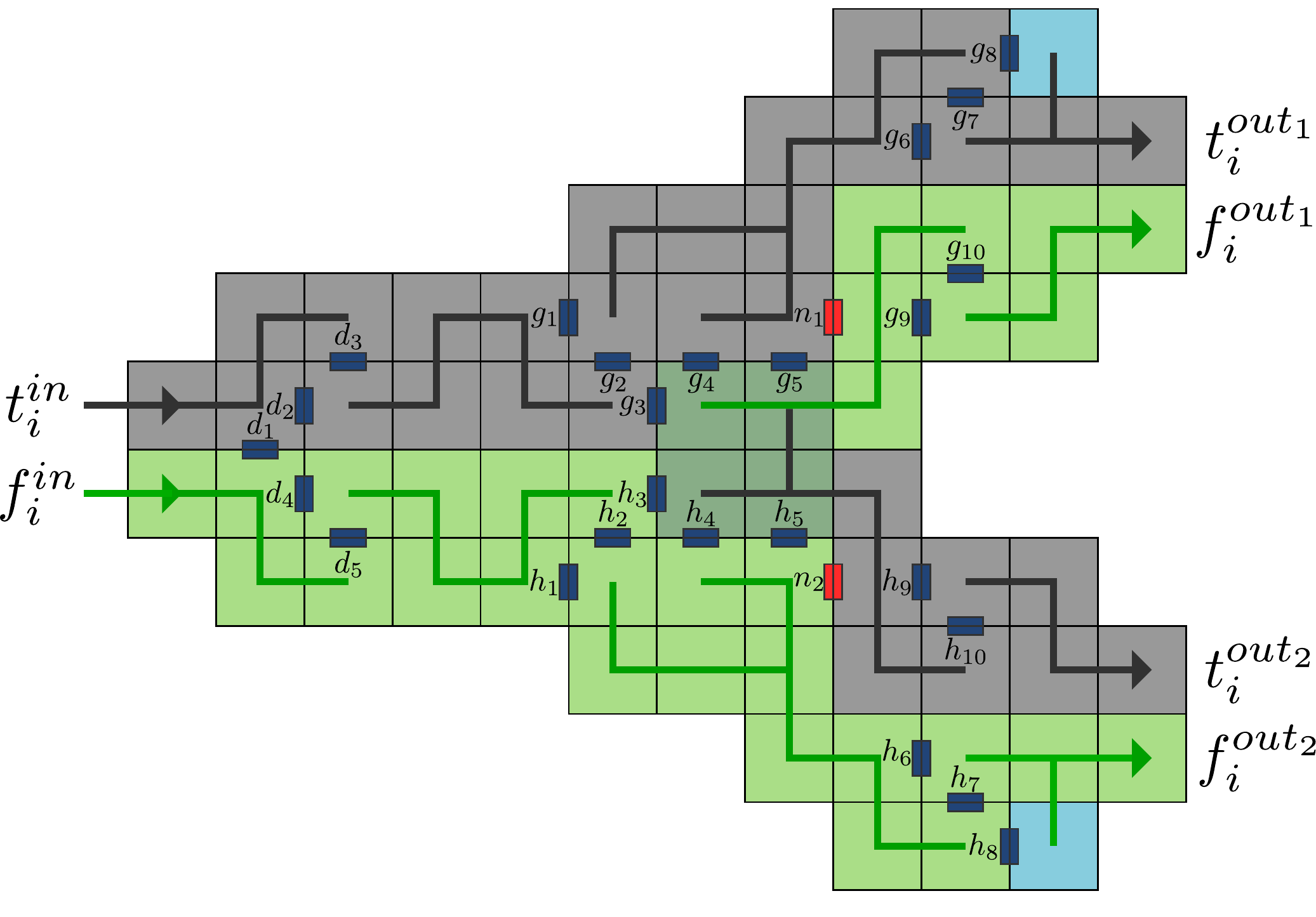}
        \caption{FANOUT}
        \label{fig:ccfanout}
    \end{subfigure}
    \begin{subfigure}[b]{0.36\textwidth}
        \includegraphics[width=1.\textwidth]{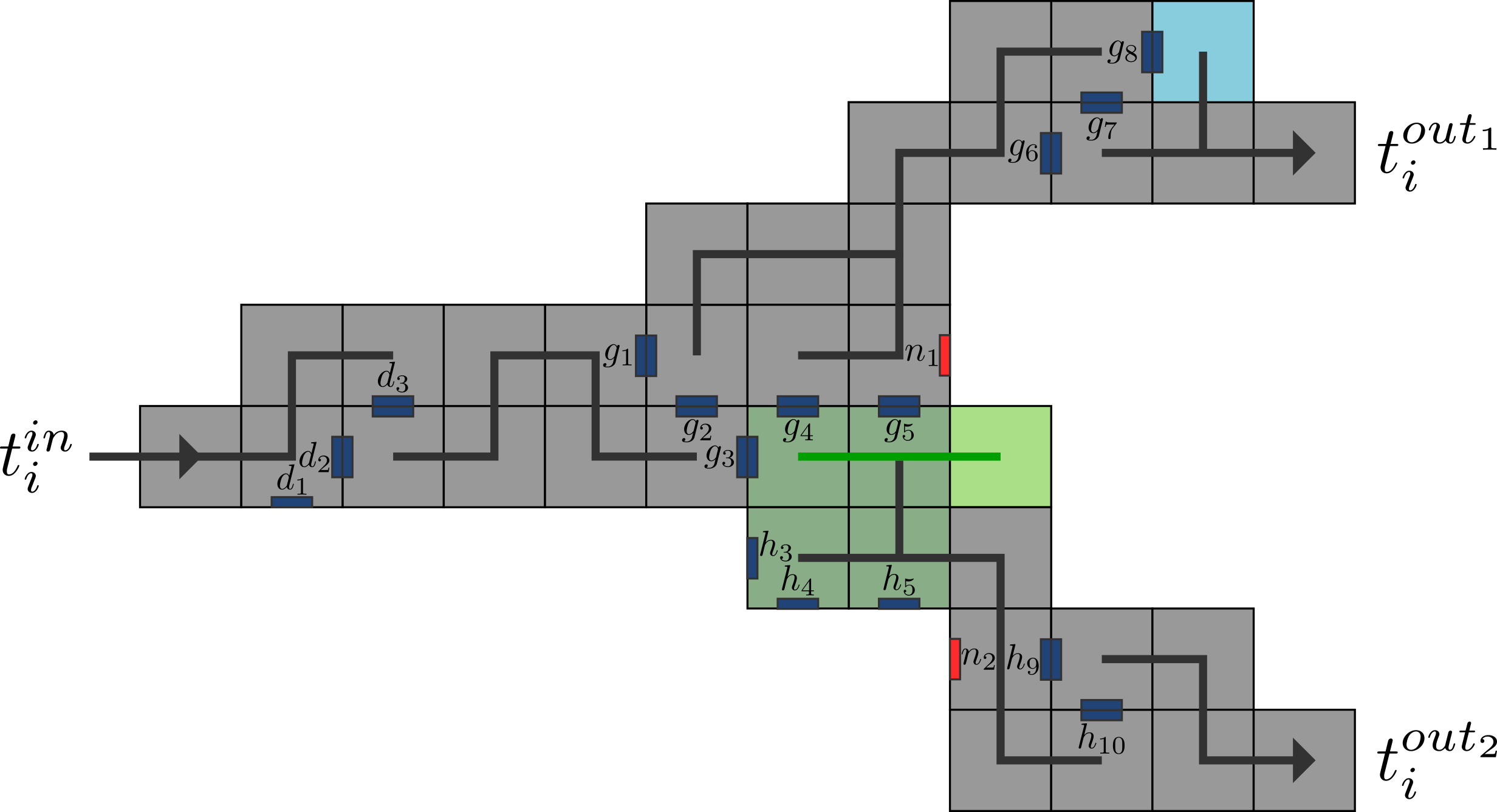}
        \caption{True Input}
        \label{fig:fots1}

        \includegraphics[width=1.\textwidth]{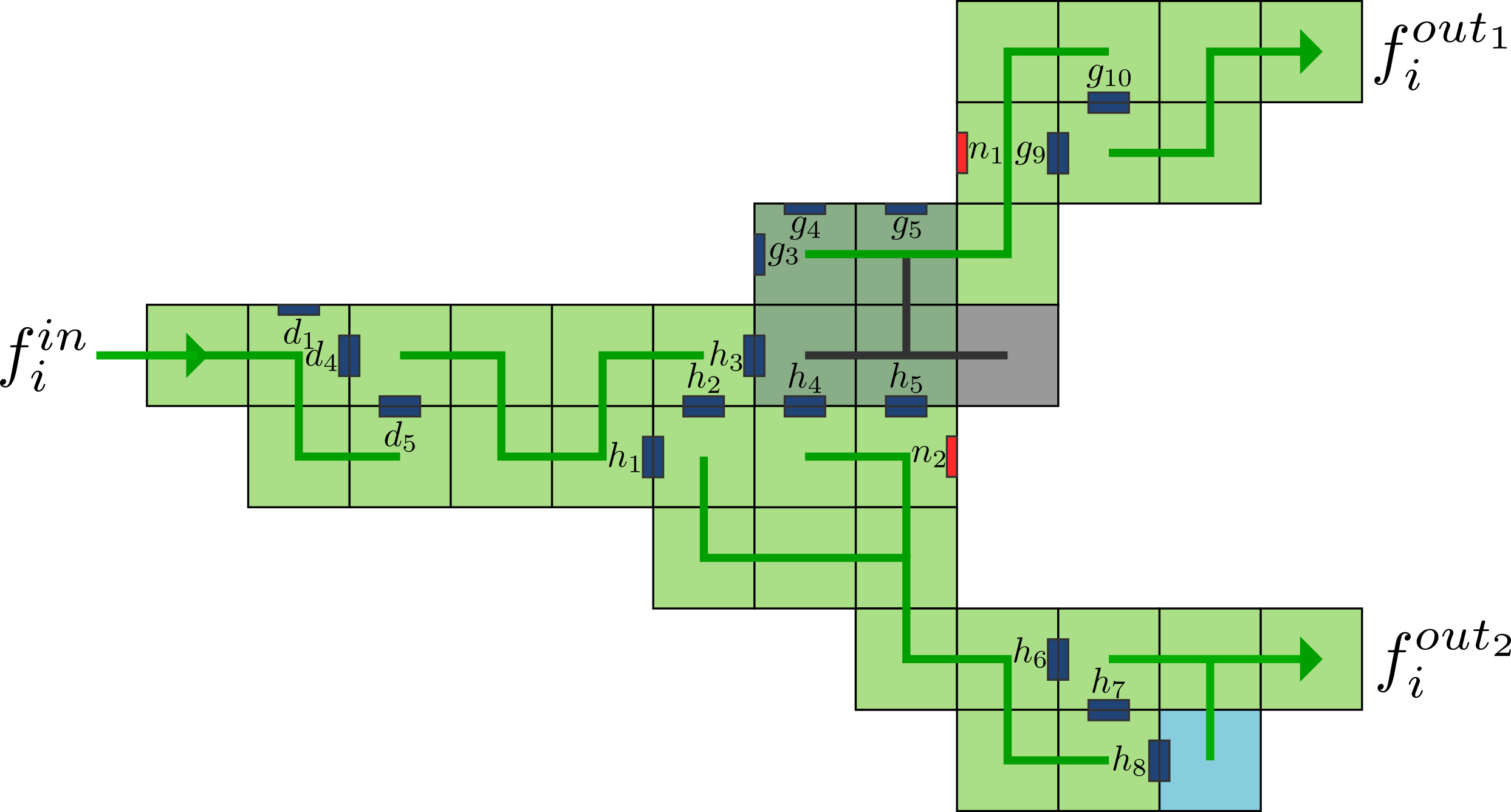}
        \caption{False Input}
        \label{fig:fofs1}
    \end{subfigure}
    %     \hspace*{.1cm}
	% \begin{subfigure}[b]{0.44\textwidth}
    %     \includegraphics[width=1.\textwidth]{images/ccORB.png}
    %     \caption{OR}
    %     \label{fig:ccorb}
    % \end{subfigure}
    %\vspace*{-.2cm}
    \caption{(a) FANOUT gadget. (b) True input wire for the FANOUT gadget $t_i^{in}$ results in output wires $t_i^{out_1}$ and $t_i^{out_2}$. (c) False input wire for the FANOUT gadget $f_i^{in}$ results in output wires $f_i^{out_1}$ and $f_i^{out_2}$.}
    \label{fig:fanor}
    %\vspace*{-.7cm}
\end{figure}

The FANOUT gadget needs to duplicate the geometric wire, and needs to backfill when at least one of the outgoing wires has backfilled. Figure \ref{fig:ccfanout} shows the FANOUT gadget. Similar to the NOT, there is a shared set of tiles placed by both the true and the false path. Figures \ref{fig:fots1} and \ref{fig:fofs1} show the true and false paths without any backfilling, respectively.  
For a FANOUT, we add a Backfill Stop gadget to both of its outputs in order to ensure they are backfilled. This is not necessary, but it ensures that the gadget always backfills even when one of the outputs is part of the output of the computation.

The FANOUT has the following necessary properties. 
%The necessary conditions decscribing the gadget are in Appendex \ref{subapp:FANOUT}.

\begin{itemize}
    \item[\textbf{1.}] With input $t_i^{in}$, the gadget outputs wires $t_i^{out_1}$ and $t_i^{out_2}$, and does not output $f_i^{out_1}$ and $f_i^{out_2}$. Figure \ref{fig:fots1} shows the true fanout without the backfilling. Due to $n_1$ and $n_2$, the false outputs can not assemble. Both settings share the same middle four tiles, but with placement order $n_1$ is placed first and then cooperative glues are used to place the first tile of the four (with glues $g_3,g_4$).

    \item[\textbf{2.}] With input $f_i^{in}$, the gadget outputs wires $f_i^{out_1}$ and $f_i^{out_2}$, and does not output $t_i^{out_1}$ and $t_i^{out_2}$. Figure \ref{fig:fofs1} shows the false fanout without the backfilling. Due to $n_1$ and $n_2$, the true outputs can not assemble.  With placement order $n_2$ is placed first and then cooperative glues are used to place the first of the four middle tiles (with glues $h_3,h_4$).

    \item[\textbf{3.}] With input wire $t_i^{in}$, wire $f_i^{in}$ only backfills once $f_i^{out_1}$ or $f_i^{out_2}$ have backfilled. Both wires backfill independently, and only $f_i^{out_2}$ can actually backfill the $f_i^{in}$ wire. However, much like the logic diodes at the input of the gadget, we require a backfill stop gadget on both of the outputs of the FANOUT. This means that both $f_i^{out_1}$ and $f_i^{out_2}$ are backfilled.

    \item[\textbf{4.}] With input wire $f_i^{in}$, wire $t_i^{in}$ only backfills once $t_i^{out_1}$ or $t_i^{out_2}$ have backfilled. Both wires backfill independently, and only $t_i^{out_1}$ can actually backfill the $t_i^{in}$ wire. However, much like the logic diodes at the input of the gadget, we require a backfill stop gadget on both of the outputs of the FANOUT. This means that both $t_i^{out_1}$ and $t_i^{out_2}$ are backfilled.  
\end{itemize}

%% file: uav.tex
%\vspace*{-.3cm}
\section{Covert Computation and Unique Assembly Verification} \label{sec:uav}
%\vspace*{-.2cm}

In this section we establish our main results related to covert computation in self-assembly systems.  We first utilize our covert circuitry to show that any function is covertly computable (Thm.~\ref{thm:covert}).  We then apply covert circuitry to show that the open problem of Unique Assembly Verification within the growth-only negative glue aTAM is coNP-complete (Thm.~\ref{thm:uav}).  %Finally, we establish that the use of negative glues is necessary for covert computation by showing that general functions are not covertly computable within the non-negative aTAM (Theorem~\ref{thm:negative}).

\begin{theorem}\label{thm:covert} For any function $f$ computed by a boolean circuit, there exists a tile assembly computer (TAC) that covertly computes $f$.
\end{theorem}

\begin{proof}
The proof of this theorem consists of a direct simulation of boolean circuits by way of a series of covert gadget implementations for various logic gates and how to connect them. 
NAND Boolean gates are functionally complete when combined with a FANOUT \cite{Sheffer:1913:TAMS}, and can implement any Boolean circuit. These are easily transformed into Circuit SAT instances to compute any function that is in the class NP \cite{Garey:1979:BOOK}, which fits within our definition of computable functions by a TAC.
Thus, the proof follows from the gadgets and machinery given in Section \ref{sec:uavgadgets} that implement the variables and gates for Circuit SAT. \hfill$\square$
\end{proof}

We now prove that Unique Assembly Verification (UAV) in a growth-only negative glue aTAM system is coNP-complete by utilizing our covert gadgets. Without the growth-only constraint, UAV in the negative glue atam is undecidable as a Turing machine simulation could use negative interactions to break down produced assemblies into a final unique terminal assembly exactly when the Turing machine halts~\cite{Doty2013}. With no negative glues however, the problem is in P \cite{ACGHKMR02}. We prove that with the ability to temporarily block, the problem becomes coNP-complete. This result is achieved with a reduction from Circuit SAT. Unique Assembly Verification in our model is formally defined as follows:

\begin{definition}[Unique Assembly Verification (growth only)]
Given a negative-glue aTAM tile-system $\Gamma=(T,S,\tau)$ with the promise that it is a growth-only system, and an assembly $A$. Does $\Gamma$ uniquely assemble $A$?
\end{definition}

%\subsection{UAV is coNP-complete}

\begin{figure}[t!]
    %\vspace*{-.2cm}
    \centering
    \begin{subfigure}[b]{0.13\textwidth}
        \centering
        \includegraphics[width=.8\textwidth]{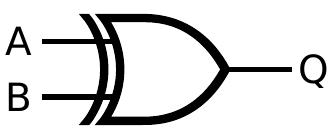}
        %\vspace*{-.2cm}
        \caption{XOR}
        \label{fig:xorsym}

        \includegraphics[width=.8\textwidth]{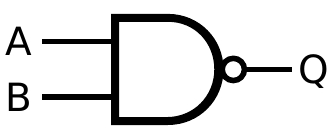}
        %\vspace*{-.2cm}
        \caption{NAND}
        \label{fig:nandsym}
    \end{subfigure}
   % \hspace*{.3cm}
	\begin{subfigure}[b]{0.17\textwidth}
	    \centering
        \includegraphics[width=.7\textwidth]{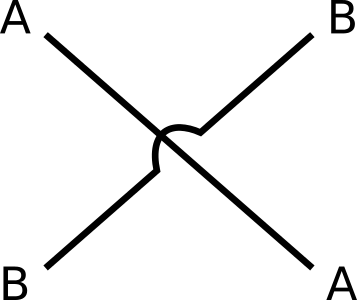}
        \caption{Crossover}
        \label{fig:cross}
    \end{subfigure}
        %\hspace*{.3cm}
	\begin{subfigure}[b]{0.29\textwidth}
	    \centering
        \includegraphics[width=1.\textwidth]{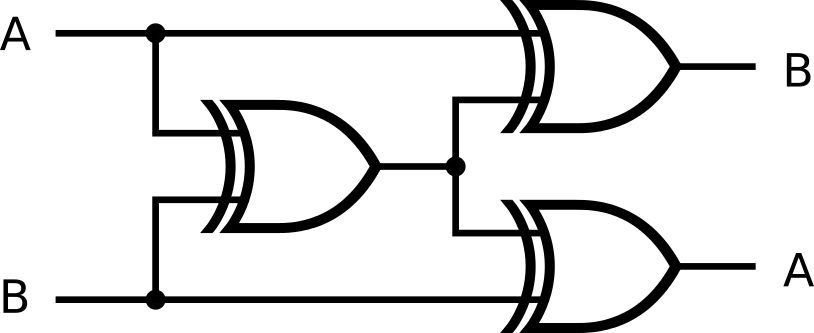}
        \caption{XOR Crossover}
        \label{fig:xorcross}
    \end{subfigure}
       %\hspace*{.3cm}
	\begin{subfigure}[b]{0.36\textwidth}
	    \centering
        \includegraphics[width=1.\textwidth]{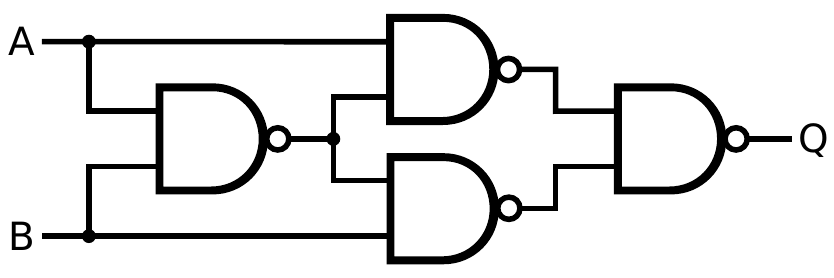}
        \caption{XOR with NANDs}
        \label{fig:xornands}
    \end{subfigure}

    %\vspace*{-.2cm}
    \caption{Constructing planar crossover gadgets with NAND gates. (a) XOR symbol. (b) NAND symbol. (c) Two wires in a circuit that cross making it non-planar. (d) A planar circuit using XOR gates that act as a crossover. (e) A planar circuit using only NAND gates that implement an XOR gate. }
    \label{fig:crossover}
    %\vspace*{-.2cm}
\end{figure}

%\textbf{Planar Circuit SAT.}
A reduction from Circuit SAT \cite{Garey:1979:BOOK} generally requires a functionally universal set of gates  and variable, wire, fanout, and crossover gadgets.  Both NAND and NOR  are functionally complete gates, so given either, all gates can be made. A crossover gadget is redundant since it can be made with XOR gates and XOR gates can be made with NAND gates \cite{Scott:2011:MI}. Figure \ref{fig:crossover} shows this derivation. Finally, Circuit SAT requires a DAG, and thus there are no cycles, and so the gadgets can be topologically sorted so that there are no crossovers that cause a loop (the output of a gadget can not crossover one of its input lines). Thus, a reduction from Planar Circuit SAT is equivalent to a reduction from Circuit SAT.

%\begin{definition}[Planar Circuit SAT]
%Instance: A planar directed acyclic graph (DAG) $G=(V,E)$ with $n$ boolean inputs, one output and every $v \in V$ is either a NAND (or NOR) gate ($deg^-(v)=2$, $deg^+(v)=1$) or a fanout ($deg^-(v)=1$, $deg^+(v)=2$). The source vertices, $v_i \in V$ s.t. $deg^-(v_i)=0$ and $1 \leq i \leq n$, are the variables. The sink vertex, $s \in V$ s.t. $deg^+(v_i)=0$ is the ``output'' of the boolean circuit.\\
%Question: Does there exist a setting of the inputs such that the output to the circuit is 1?
%\end{definition}

\begin{definition}[Planar Circuit SAT]
Given a planar directed acyclic graph (DAG) $G=(V,E)$ with $n$ boolean inputs, one output, and every $v \in V$ is either a NAND (or NOR) gate ($deg^-(v)=2$, $deg^+(v)=1$) or a fanout ($deg^-(v)=1$, $deg^+(v)=2$), the source vertices ($v_i \in V$ s.t. $deg^-(v_i)=0$ and $1 \leq i \leq n$) are the variables, and the sink vertex ($s \in V$ s.t. $deg^+(v_i)=0$) is the ``output'' of the boolean circuit.
Does there exist a setting of the inputs such that the output to the circuit is 1?
\end{definition}

\begin{theorem} \label{thm:uav}
  Unique Assembly Verification in the aTAM with repulsive forces in a growth only system is coNP-complete.
\end{theorem}

\begin{proof}
We first observe that Unique Assembly Verification with repulsive forces is in coNP as any failure to uniquely assemble a target assembly $A$ comes in the form of a polynomially sized assembly that is inconsistent with $A$.  The producibility of this assembly can be verified in polynomial time, and thus serves as a certificate for ``no'' instances to the UAV problem.

We now show coNP-hardness by a reduction from Planar Circuit SAT.  Assume we are given an arbitrary instance of planar Circuit SAT $C$ with inputs $i_1, \dots, i_n$ where $i\in\{0,1\}$, i.e., a boolean circuit. By our definition we assume there are only NAND gates, fanouts, input variables and an ouput variable in the planar DAG. %we can transform every gate into a functionally equivalent circuit, in linear time, to a boolean circuit that consists of only NAND gates, fanouts, and wires, and thus we assume each gate is a NAND gate. Since this is a boolean circuit we topologically order the gates from input to output to ensure proper planarity and processing of the gates.

For our reduction, we build a tileset $T$ by adding tiles corresponding to the covert gadgets and connections described in Section~\ref{sec:uavgadgets}. Replace each NAND gate with a unique set of tiles implementing a NAND gadget, and each FANOUT gate with a unique set of tiles implementing a FANOUT gadget. For each edge, a unique sequence of tiles is added to $T$ that connects the two gadgets representing the two gates the edge connected.

This yields a tile assembly computer (TAC), $\Im=(T, I, O, \tau)$, for covertly computing the circuit $C$.  The key modification to show coNP-hardness is the utilization of a seed that non-deterministically grows any one of the possible $n$-bit input seeds for this TAC, and then evaluates the circuit.  If the circuit is not-satisfiable, then the final computation will be false regardless of the guessed input, and therefore will yield the unique ``no'' assembly of the TAC based on the fact that the circuit is computed covertly.  On the other hand, if there exists some satisfying $n$-bit input, there will be at least one final assembly that differs from the ``no'' assembly.  Thus, the ``no'' assembly is uniquely produced if and only if the circuit $C$ is not satisfiable, thereby showing coNP-hardness.

\emph{Non-deterministic input selection.}  To non-deterministically form the possible input bits, we include the tile types and seed tile described in Figure~\ref{fig:ccvar}.  The seed grows a length $O(n)$ line with each bit being encoded by a pair of adjacent locations which expose a glue on the north edge.  For each pair of positions, the presence of the left tile denotes a ``1'' for the respective bit, and the placement of the right tile denotes a ``0''.  The ``1'' and ``0'' tiles share a negative strength $1$ glue, making their mutual placement impossible until the covert gadgets have passed on the computed signal and backfilled. \hfill$\square$
%From this we replace each NAND gate with a covert NAND gadget as described, and connect the gadgets with wires as given. Any junction of wires is replaced by a FANOUT gadget (or multiple ones). We then construct a big seed $S$ consisting of variables with the appropriate setting for each. This seed will begin the self-assembly process. There are two possible terminal assemblies $A_0,A_1$ of our designed circuit.
%%see construction for reduction.
%
%Circuit $C$ will output $1$ if and only if tile system $\mathcal{T}=(T,S,...,\tau=2)$ with seed $S$ self-assembles $A_1$.
%
%Forward Direction.
%
%Backward Direction
%
%This is the shittiest hand-waving writeup ever with no formalization at all.
%
%membership in coNP.
\end{proof}

Given that UAV is coNP-complete with negative glues by way of covert circuitry, yet UAV is in $P$ without negative glues~\cite{ACGHKMR02}, it is reasonable to conjecture that the use of negative interactions is needed to perform covert computation.  %For our final result, we formally prove that general covert computation is not possible without negative glues.

% \begin{conjecture}
% For some function $f$ computed by a boolean circuit, there does not exist a tile assembly computer (TAC) that covertly computes $f$ in the aTAM without negative glues.
% \end{conjecture}

% For our final result, we show that covert computation in the aTAM is not possible without repulsive forces.  The ability to block makes UAV coNP-complete, but UAV is polynomial in the aTAM without blocking \cite{ACGHKMR02}. Here we show that covert computation in the aTAM without repulsive forces is not possible. More formally,

\begin{conjecture}\label{thm:negative} For some function $f$ computed by a boolean circuit, there does not exist a polynomially-sized tile assembly computer (TAC) that covertly computes $f$ in the aTAM without negative glues.
\end{conjecture}

%% file: examples.tex
\section{Further Motivation} \label{sec:fm}
Here, we give a few more motivating examples and some simplified gadgets. There is a lot of future work in this vein of research that is extremely relevant to modern society. We first cover the covert AND and OR gadgets.

\subsection{Simplified Gadgets} \label{subsec:simpgad}
Even though NAND gates alone are functionally complete, for some gates the circuit is larger than desired. Here, we give compact direct versions of some other useful gadgets and gates. This does not affect the complexity, but does help build a more efficient covert computation toolkit.

\begin{figure}[t!]
    \centering
	\begin{subfigure}[b]{.48\textwidth}
        \includegraphics[width=1.\textwidth]{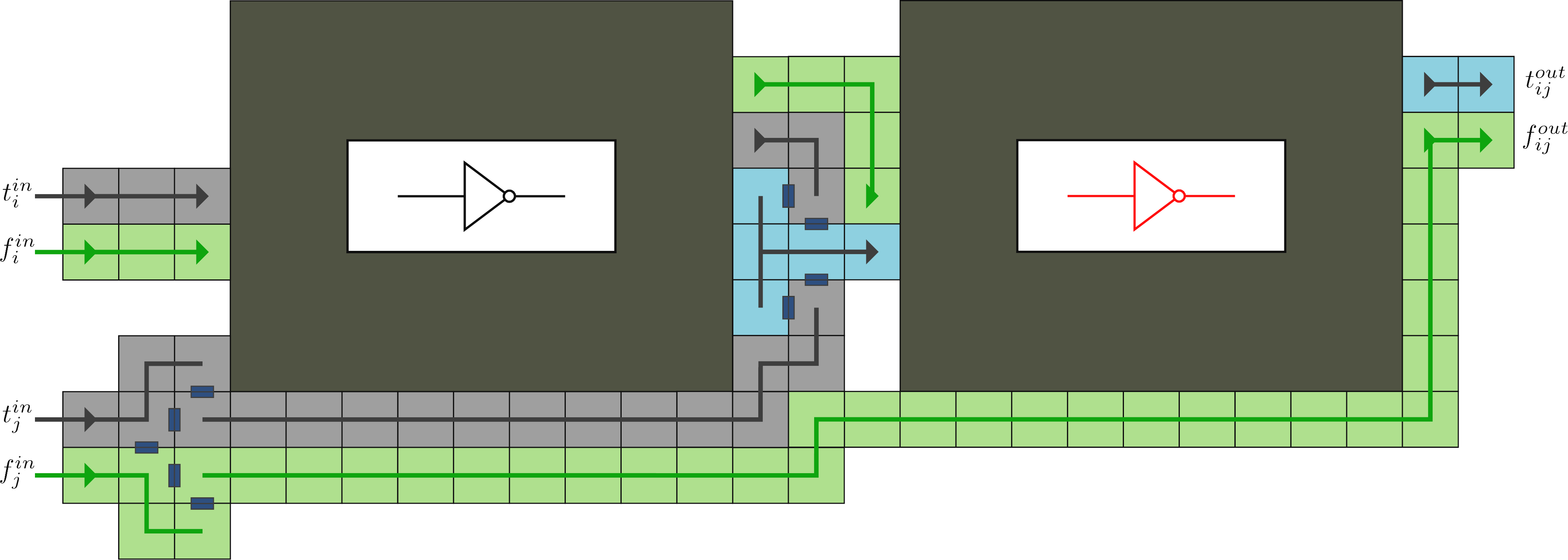}
        \caption{AND Block Diagram}
        \label{fig:ccandb}
    \end{subfigure}
        %\hspace*{.1cm}
	\begin{subfigure}[b]{.48\textwidth}
        \includegraphics[width=1.\textwidth]{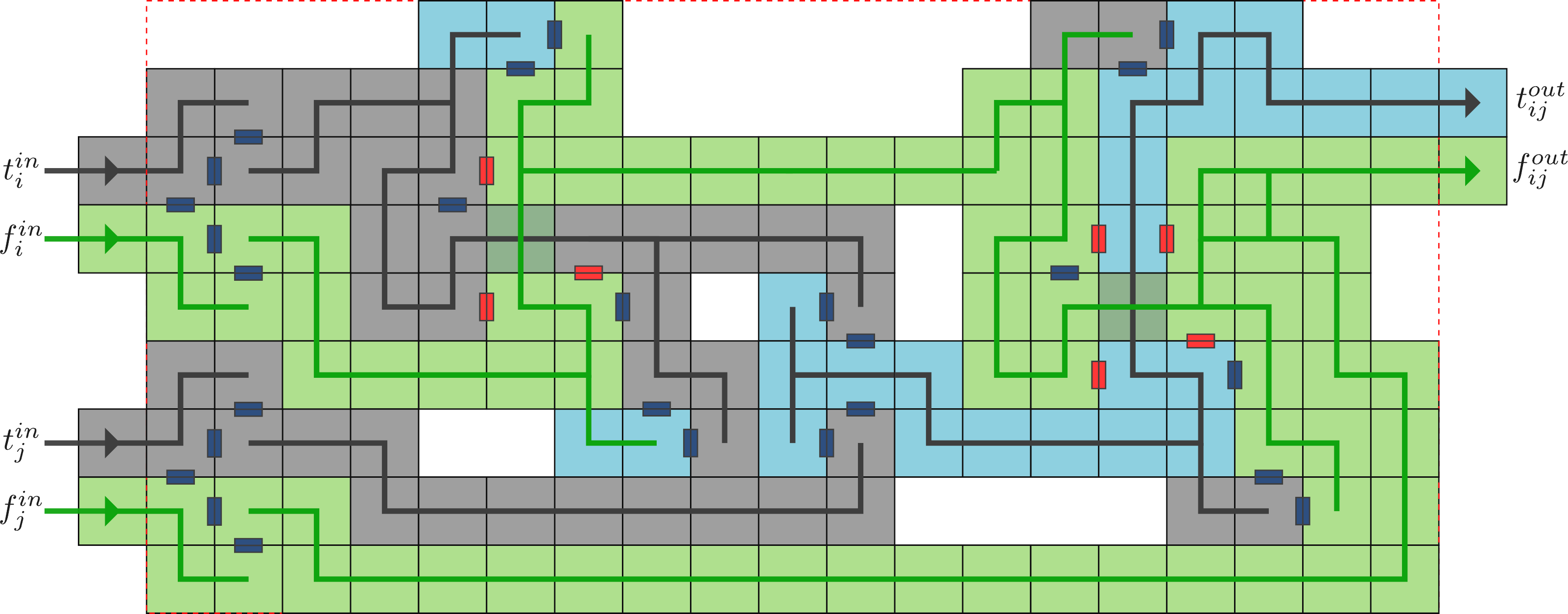}
        \caption{AND}
        \label{fig:ccand}
    \end{subfigure}
    %\vspace*{-.7cm}
    \caption{(a) Diagram of the covert AND gate with NOTs shown as blocks. The left box is the standard NOT gadget and the right box is the V-NOT gadget (has an additional vertical glue). (b) The full AND gate with the two NOT gadgets filled in and some simplification for space.}
    \label{fig:andgadget}
    %\vspace*{-.8cm}
\end{figure}

\textbf{Covert AND Gadget.}
The covert AND gadget is nearly identical to the NAND gadget. The only real difference is which two inputs the second NOT takes in. Also, similar to the H-NOT needed for the NAND, we create a V-NOT, which is a NOT with one additional vertically aligned negative glue. Figure \ref{fig:ccandb} shows the AND gadget with the blocks in place of NOTs for clarity, and Figure \ref{fig:ccand} shows the full gadget.

\begin{figure}[ht!]
    \centering
	\begin{subfigure}[b]{0.48\textwidth}
         \includegraphics[width=1.\textwidth]{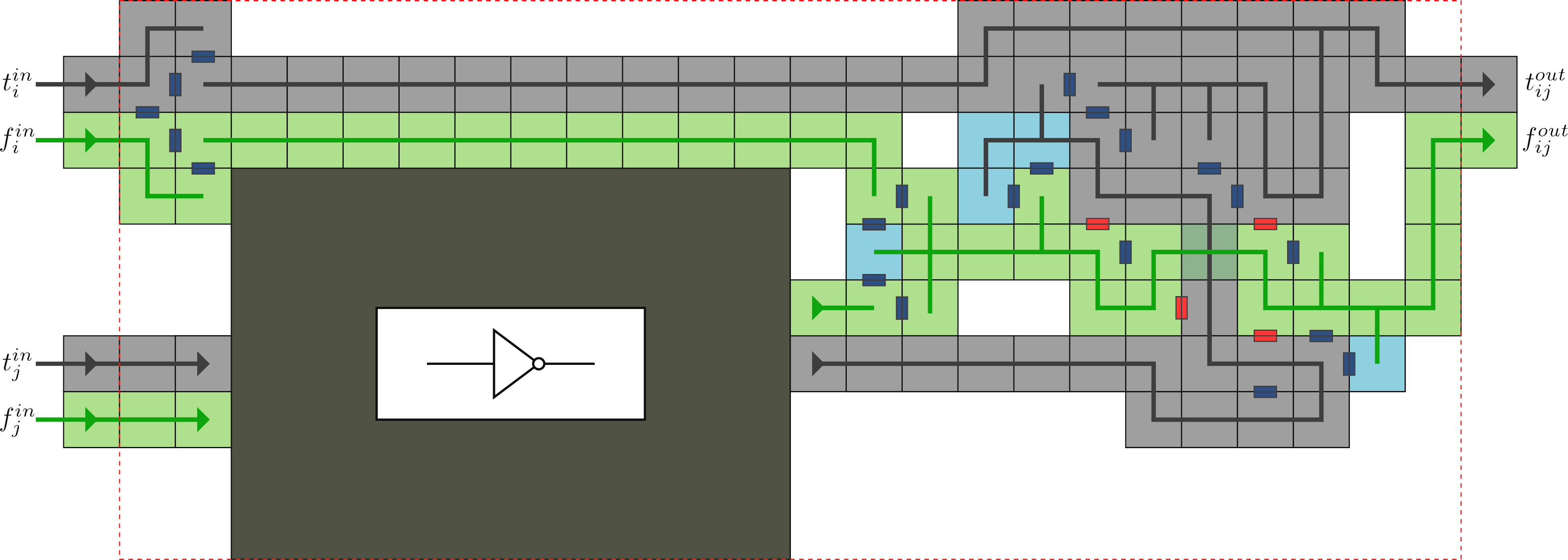}
         \caption{Block OR}
         \label{fig:ccorb}
     \end{subfigure}
     \begin{subfigure}[b]{0.48\textwidth}
         \includegraphics[width=1.\textwidth]{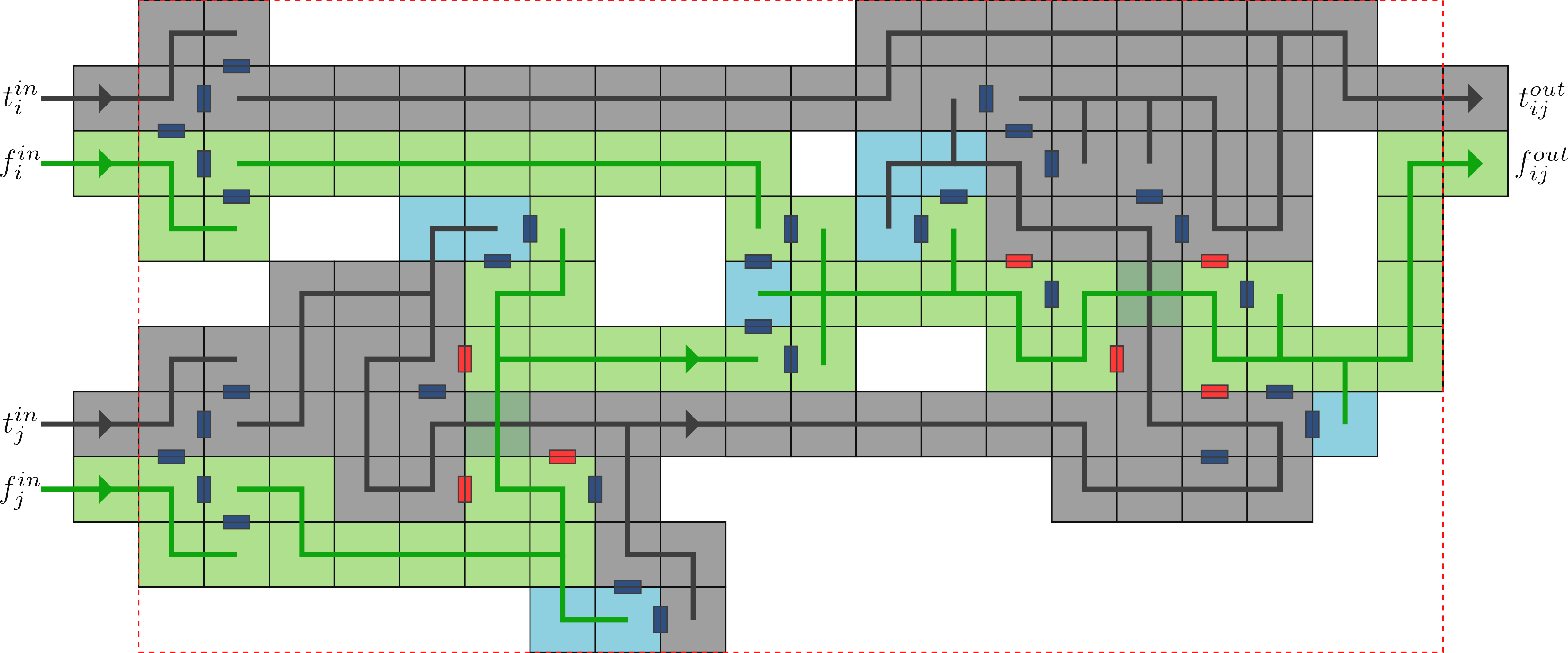}
         \caption{OR}
         \label{fig:ccor}
     \end{subfigure}
    \caption{(a) Block diagram for the OR gadget. (b) The covert OR gadget with the NOT gadget block filled in with the tiles.}
    \label{fig:ors}
    %\vspace*{-.8cm}
\end{figure}

\textbf{Covert OR Gadget.}
The covert OR gadget still uses a NOT to flip one of the inputs, but does several checks on the second flip to the point of drastically differing from a NOT. This is similar to needing H-NOT and V-NOT in previous gadgets to handle specific issues related to the boolean function, but more changes are required. Figure \ref{fig:ccorb} shows the OR gadget with the blocks in place of NOTs for clarity, and Figure \ref{fig:ccor} shows the full gadget.

\subsection{Encryption and Cryptography} \label{subsec:crypt}
\begin{figure}[t!]
    %\vspace*{-.2cm}
    \centering
    \begin{subfigure}[b]{0.15\textwidth}
    \centering
        \includegraphics[width=.9\textwidth]{XOR_ANSI_Labelled.pdf}
        \caption{XOR}
        \label{fig:xorsym2}

        \includegraphics[width=.9\textwidth]{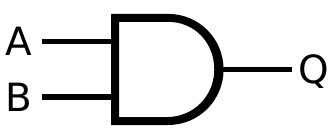}
        \caption{AND}
        \label{fig:andsym}
    \end{subfigure}
    \hspace*{.3cm}
	\begin{subfigure}[b]{0.27\textwidth}
	\centering
        \includegraphics[width=1.\textwidth]{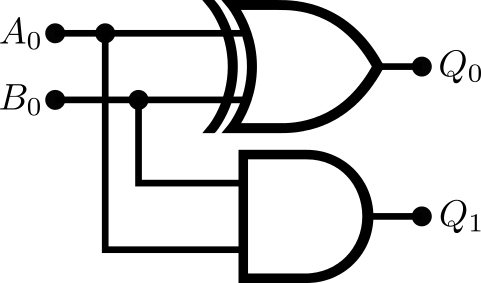}
        \caption{Half-Adder}
        \label{fig:halfadder}
    \end{subfigure}
    \hspace*{.3cm}
	\begin{subfigure}[b]{0.42\textwidth}
	\centering
        \includegraphics[width=.8\textwidth]{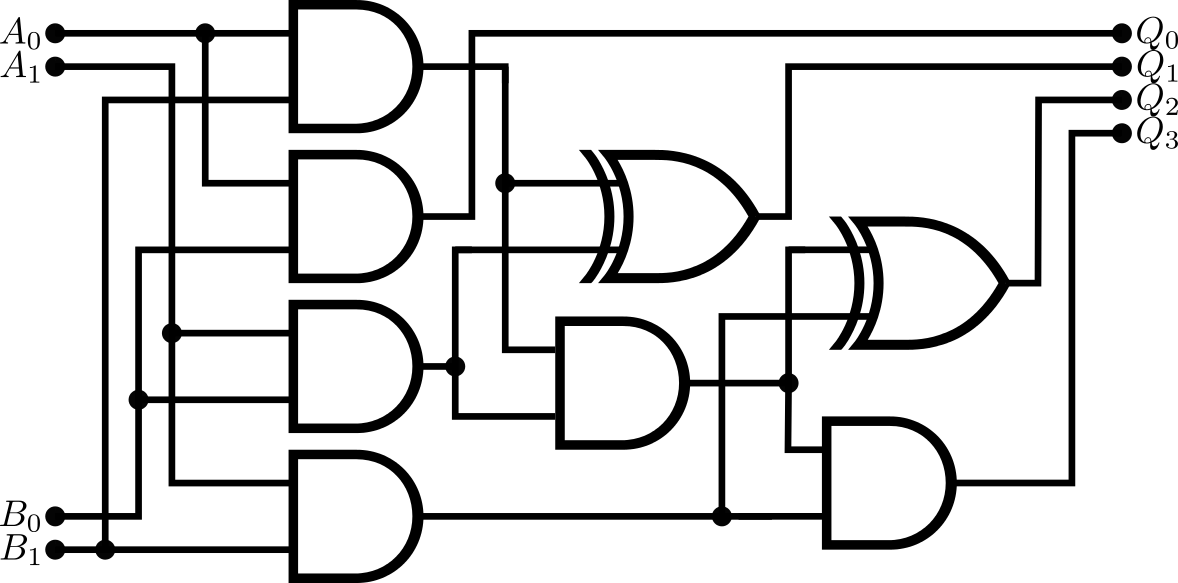}
        \caption{2-Bit Multiplier}
        \label{fig:2bitmult}
    \end{subfigure}

    %\vspace*{-.2cm}
    \caption{Constructing covert circuits for arithmetic building up to cryptography examples. (a) XOR symbol. (b) AND symbol.  (c) A half-adder, which has two 1-bit numbers as input and a 2-bit number as output. (d) A 2-bit multiplier which has two 2-bit numbers as input and outputs a 4-bit number that is their product. This can be expanded to use two large primes resulting in a large number that would be hard to factor.}
    \label{fig:mult}
    %\vspace*{-.2cm}
\end{figure}

Several encryption methods are based off problems that we believe to be ``hard'' computationally. One of the most common is factoring the product of large prime numbers, which is the basis for several encryption schemes. %methods such as RSA \cite{}, etc....
Although factoring may be difficult, the function to generate the number is simple multiplication, which can be accomplished with simple circuits. Figure \ref{fig:2bitmult} shows a simple 6-gate circuit implementing a 2-bit number multiplier resulting in a 4-bit output number. An $n$-bit multiplier scales linearly (in the number of bits) with additional AND gates and full and half adders.

\begin{figure}[t!]
    %\vspace*{-.2cm}
    \centering
    \begin{subfigure}[b]{0.1\textwidth}
	\centering
        \includegraphics[width=1.\textwidth]{ccHAseed.pdf}
        \caption{Seed}
        \label{fig:cchaseed}
    \end{subfigure}
    \begin{subfigure}[b]{0.87\textwidth}
        \centering
        \includegraphics[width=1.\textwidth]{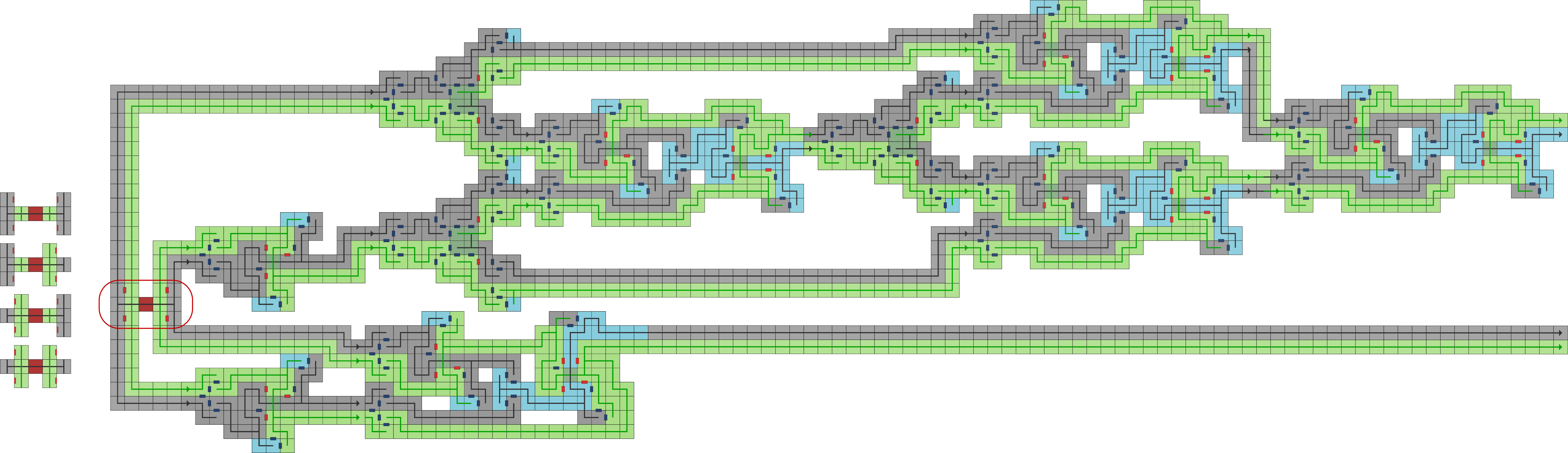}
        \caption{Covert Half-Adder}
        \label{fig:cchalfadder}
    \end{subfigure}
    %\vspace*{-.2cm}
    \caption{Covert Half-Adder made with 4 NANDs, 3 FANOUTs, 2 NOTs, and 1 AND. The seed input is highlighted and all 4 possible seeds are shown in (a). Regardless of the seed, the final assembly will look identical except the final T/F representing the bits of the numbers added. This implements the schematic shown in Figure \ref{fig:halfadder} and the XOR is implemented with NANDs as shown in Figure \ref{fig:xornands}.}
    \label{fig:ccha}
    %\vspace*{-.2cm}
\end{figure}

Implementing the multiplier with covert gates is not difficult, but the resulting assembly is large due to the inefficient crossover gadget used. Instead, we demonstrate a simple half-adder. The schematic for a half-adder is in Figure \ref{fig:halfadder}. A covert half-adder as a TAC is shown in Figure \ref{fig:cchalfadder}. The XOR has been replaced by the 4 NAND gates as shown in Figure \ref{fig:xornands}. Further, 3 FANOUTs were needed, an AND gadget as shown above in Section \ref{subsec:simpgad}, and 2 NOT gadgets were used to flip the input for the gadgets. Figure \ref{fig:cchaseed} shows the four possible input seeds to build the assembly. %A half-adder is simple enough to know which seed was used if $00$ or $10$ are output, but if $01$ is output there is no way to know.

%\subsection{Biomedical}
%aoeu

%\subsection{Intellectual Property}
%military, genetically modified crops, etc.

%% file: conclusion.tex
\section{Conclusions and Future Work}\label{sec:conc}
We have introduced the concept of covert computation in self-assembly and provided a general scheme to implement any boolean circuit under this restriction.  Beyond potential applications to biomedical privacy, cryptography, and intellectual property, our techniques and framework promise to impact self-assembly theory itself.  As a first example we have applied our techniques to the fundamental problem of Unique Assembly Verification in the negative glue aTAM, and shown it to be coNP-complete with growth-only systems, essentially as a corollary of our covert computation theory.

A number of future directions stem from our work.  Having established the general computation power of covert computation, a natural next step is the consideration of efficiency for computing classes of functions.  The \emph{time} complexity of self-assembly computation has been studied~\cite{BRUN2007,Keenan2016} and shown to allow for a substantial amount of parallelism.  Can similar results be achieved under the covert constraint?  What general connections exists between the time complexity for unrestricted self-assembly computation versus that of covert computation?  Other natural metrics include minimizing the number of distinct tile types, along with the space taken up by the final assembly of the computation.  %One simple initial question is what is the simplest, or smallest, possible covert NAND gate?

Another future direction is to investigate what model features are required for covert computation. In this paper, we conjectured that there does not exists a tile assembly computer that performs covert computation in the aTAM without negative glues. This conjecture is made with the assumption that no other model definitions change. If other definition relaxations are made to the model, such as allowing assemblies which are equal up to translation to be considered the same assembly, might covert computation be possible without negative glues?